\documentclass[12pt,preprint]{aastex}
\usepackage{graphicx}
\usepackage{txfonts}
\usepackage{latexsym}
\usepackage{amssymb}
\usepackage[activeacute,english]{babel}
\usepackage{natbib}
\begin{document}

\title{WAVE LEAKAGE AND RESONANT ABSORPTION IN A LOOP EMBEDDED IN A CORONAL ARCADE}
\shorttitle{Wave leakage and resonant absorption in a coronal loop embedded in a coronal arcade.}
\author{S. Rial, I. Arregui, J. Terradas, R. Oliver, J. L. Ballester}

   \affil{Departament de F\'isica, Universitat de les Illes Balears,
              E-07122, Palma de Mallorca, Spain}
              \email{[samuel.rial;inigo.arregui;jaume.terradas;ramon.oliver;joseluis.ballester]@uib.es}
\date{ }

\begin{abstract}

We investigate the temporal evolution of impulsively generated perturbations in a potential coronal arcade with an embedded loop. As the initial configuration we consider a coronal loop, represented by a density enhancement, which is unbounded in the ignorable direction of the arcade. The linearized time-dependent magnetohydrodynamic equations have been numerically solved in field-aligned coordinates and the time evolution of the initial perturbations has been studied in the zero-$\beta$ approximation. For propagation constrained to the plane of the arcade, the considered initial perturbations do not excite trapped modes of the system. This weakness of the model is overcome by the inclusion of wave propagation in the ignorable direction. The inclusion of perpendicular propagation produces two main results. First, damping by wave leakage is less efficient because the loop is able to act as a wave trap of vertical oscillations. Second, the consideration of an inhomogeneous corona enables the resonant damping of vertical oscillations and the energy transfer from the interior of the loop to the external coronal medium. 
\end{abstract}


\section{Introduction}
\label{Introduction}

In the last years a phenomenon that has attracted the attention of solar physicists is the discovery of transverse oscillations of coronal loops observed in EUV wavelengths ($171$ \AA) with TRACE in $1998$. The oscillatory amplitude and period are of the order of a few Mm and a few min, respectively, and one of the most interesting features of these oscillations is that their amplitude decreases quickly with time, typically in a few periods. Some examples of this oscillatory phenomenon, namely flare-excited transversal oscillations, were reported by \citet{AFS1999,NOD1999}, more recently by \citet{WS2004,HIS2005}, and \citet{AS2011}, see \citet{APS2002,SAT2002} and \citet{A2009} for an extensive overview and analysis of transversal flare-excited coronal loop oscillations. 

The first theoretical studies of the modes of oscillation of coronal flux tubes modeled as straight magnetic cylinders were done by  \citet{W1979}, \citet{S1981b}, \citet{ER1983} and \citet{REB1984}. Later, flare-generated transverse oscillations were interpreted as fast kink eigenmode oscillations of straight cylindrical tubes \citep{NO2001,RR2002,GAA2002}. In the context of curved coronal magnetic structures, \citet{GPH1985,PHG1985,PG1988} investigated the continuous spectrum of ideal magnetohydrodynamics (MHD). \citet{OBH1993,OHP1996} and \citet{TOB1999} derived the spectrum of modes in potential and non-potential arcades. More complex configurations, such as sheared magnetic arcades in the zero plasma-$\beta$ limit, have been studied by \citet{AOB2004,AOB2004a}. Other authors have studied eigenmodes in curved configurations with density enhancements that represent coronal loops \citep[e.g.,][]{VDA2004,TOB2006a,VFN2006a,VFN2006b,VFN2006c,DZR2006,VVT2009}. The fact that the corona is a highly inhomogeneous and structured medium complicates the theoretical description of MHD waves and it is believed that this could be the underlying cause of the observed wave damping. Several mechanisms of wave damping have been proposed, the most popular being phase mixing \citep{HP1983}, resonant absorption \citep[and references therein]{HY1988,G1991,GAA2002,RR2002,VDA2004} and more recently wave leakage by tunneling \citep{BA2005,BVA2006,VFN2006a,VFN2006b,VFN2006c,DZR2006}.

Although normal modes should be seen as the building blocks to interpret coronal loop oscillations they do not represent the whole picture, but their study provides a basis for understanding the dynamics of the system. To have a more accurate description, the time-dependent problem needs to be analyzed. Using this method, \citet{CB1995,CB1995a} studied analytically the propagation of fast waves in a two-dimensional coronal arcade with uniform Alfv\'en speed. \citet{OMB1998} studied the effect of impulsively generated fast waves in the same coronal structure. \citet{DSV2005} studied the properties of Alfv\'en waves in an arcade configuration, including a transition region between the photosphere and the corona. \citet{TOB2008} used a potential arcade embedded in a low $\beta$ environment to study the properties of linear waves. Other studies have analyzed the effect of the loop structure on the properties of fast and slow waves in two-dimensional curved configurations \citep[see, e.g.,][]{BA2005,MSN2005,BVA2006,SSM2006,SMS2007}; see \citet{T2009} for a review.

The aim of this work is to examine two physical mechanisms involved in the fast attenuation of the observed vertical coronal loop oscillations. Previous studies by \citet{VFN2006a,VFN2006b,VFN2006c,SSM2006,SMS2007} have considered the damping of vertical oscillations by wave leakage in a coronal loop embedded in a curved configuration. When propagation is constrained to the plane of the structure, these authors find that wave leakage produces a strong damping and that the ratio of the damping time to the period is, in some cases, much shorter than the observed values. On the other hand, damping by resonant absorption, caused by the inhomogeneity of the medium, has mainly been studied in single magnetic slabs \citep{TOB2005a,ATO2007} and in single magnetic cylinders \citep{RR2002,TOB2006}, for example. Nevertheless, more complex equilibrium models have also been considered \citep{VDA2004,TOB2006a,TAO2008}. \citet{RAT2010} have recently considered the coupling of fast and Alfv\'en modes in a potential coronal arcade with three-dimensional propagation of perturbations. This study shows that because of the inclusion of perpendicular propagation fast wave energy can easily be converted into Alfv\'en wave energy at given magnetic surfaces by means of resonant coupling. Our analysis aims at extending the model by \citet{RAT2010} by including a density enhancement in a curved magnetic configuration in order to study how three-dimensional propagation affects the efficiency of the damping of vertical loop oscillations by wave leakage and how the inhomogeneity of the corona can produce coupling of modes and energy transfer.

The paper is arranged as follows. In \S~\ref{equilibrium_conf} we describe the equilibrium configuration as well as the approximations made in this work. In \S~\ref{eq_linear} we present the linear ideal MHD wave equations with three-dimensional propagation of perturbations. In \S~\ref{method} we describe the numerical setup together with the initial and boundary conditions. Our results are shown in \S~\ref{results}, where the linear wave propagation properties of coupled fast and Alfv\'en waves in a two-dimensional coronal loop, allowing three-dimensional propagation, are described. Finally, in \S~\ref{conclusions} the conclusions are drawn.

\section{Equilibrium configuration}  
\label{equilibrium_conf}

\begin{figure}[!h]
\begin{center}
\includegraphics[width=0.49\textwidth]{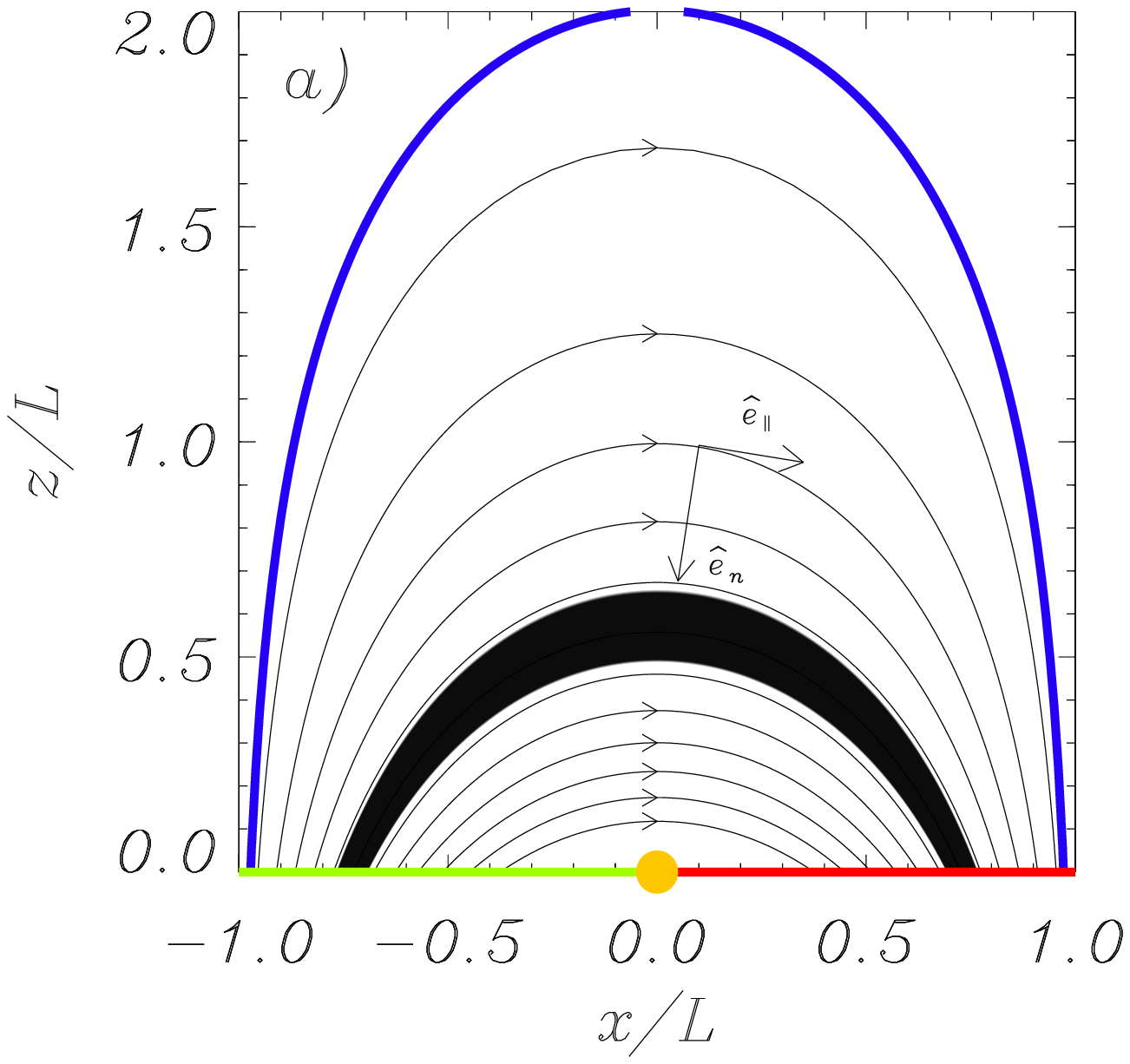}
\includegraphics[width=0.49\textwidth]{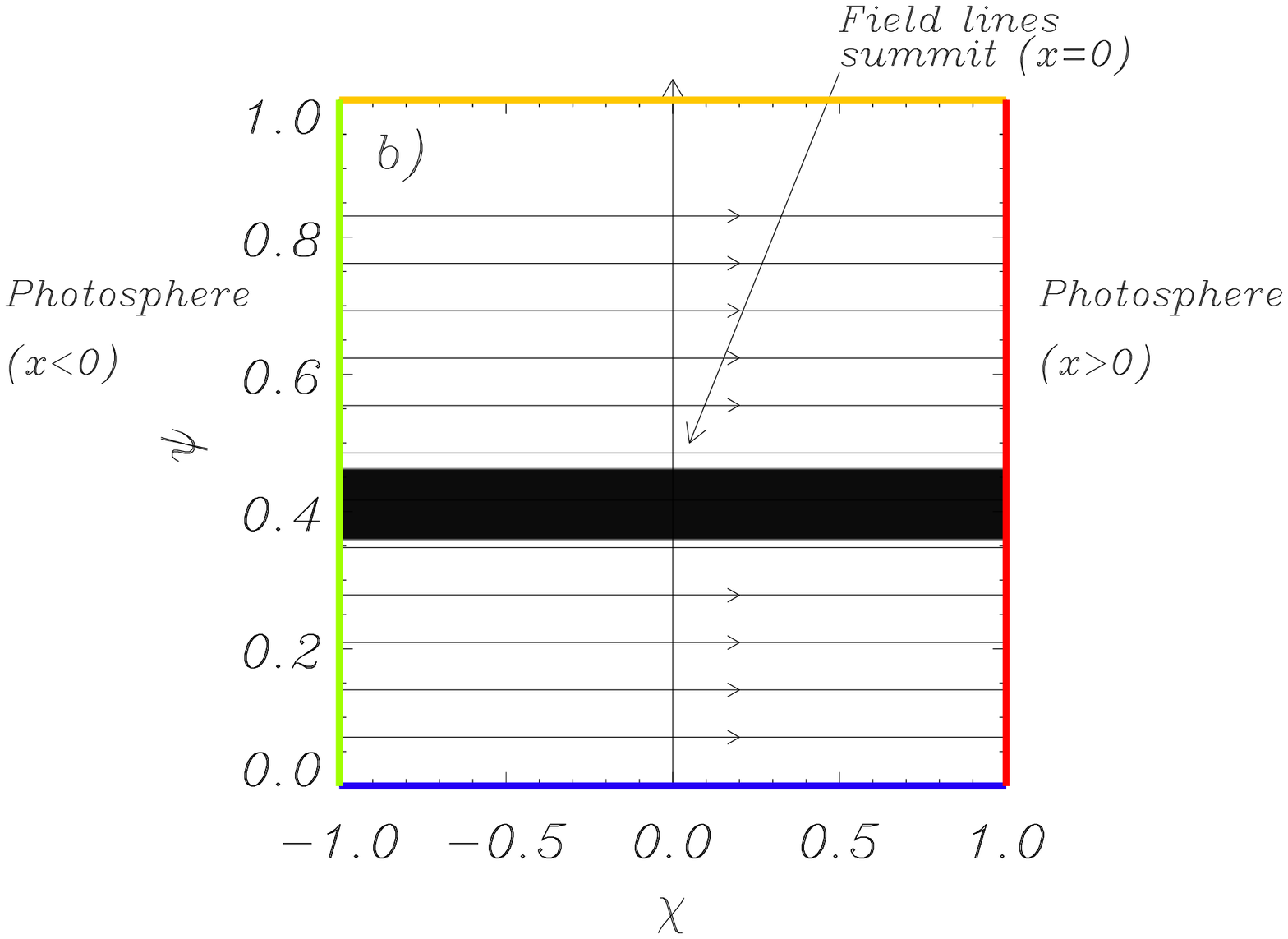}
\caption{(a) Magnetostatic configuration of the potential coronal arcade described by Equations~(\ref{eq:flux}) and (\ref{eq:arccomp}), where the solid curves represent magnetic field lines, given by $A(x,z) = $constant. These curves in the $xz$-plane become arcade surfaces in three dimensions. In this model $z$ measures the vertical distance from the base of the corona (placed at $z=0$). The black region represents a coronal loop embedded in the arcade. Two orthogonal unit vectors defining the normal and the parallel directions, $\hat{e}_{n}$ and  $\hat{e}_{\Vert}$, are also shown at a particular point. (b) Same as (a) in the $\chi\psi$-plane. Curved magnetic field lines in the $xz$-plane become straight lines in this flux coordinate system and therefore a curved loop becomes a straight slab. The four boundaries are color-coded to show their correspondence with lines and points in panel (a)}
\label{fig:density}
\end{center}
\end{figure}

The equilibrium magnetic field is a potential arcade contained in the $xz$-plane \citep[see][for more details]{OBH1993}. In Cartesian coordinates the flux function is

\begin{equation} 
A(x,z)=B_{0}\Lambda_{B}\cos{\left(\frac{x}{\Lambda_{B}}\right)}\exp{\left(-\frac{z}{\Lambda_{B}}\right)},
\label{eq:flux}
\end{equation}
 and the magnetic field components are given by

\begin{displaymath} 
B_{x}(x,z)=B_{0}\cos\left(\frac{x}{\Lambda_{B}}\right)\exp\left({-\frac{z}{\Lambda_{B}}}\right),
\end{displaymath}
\begin{equation}
B_{z}(x,z)=-B_{0}\sin\left(\frac{x}{\Lambda_{B}}\right)\exp\left({-\frac{z}{\Lambda_{B}}}\right).
\label{eq:arccomp}
\end{equation}
In these expressions $\Lambda_{B}$ is the magnetic scale height, which is related to the lateral extent of the arcade, $2L$, by $\Lambda_{B}=2L/\pi$, and $B_{0}$ is the magnetic field strength at the base of the corona ($z=0$). The overall shape of the arcade is shown in Figure~\ref{fig:density}a. 

\begin{figure}[h]
\begin{center}
\includegraphics[width=0.5\textwidth]{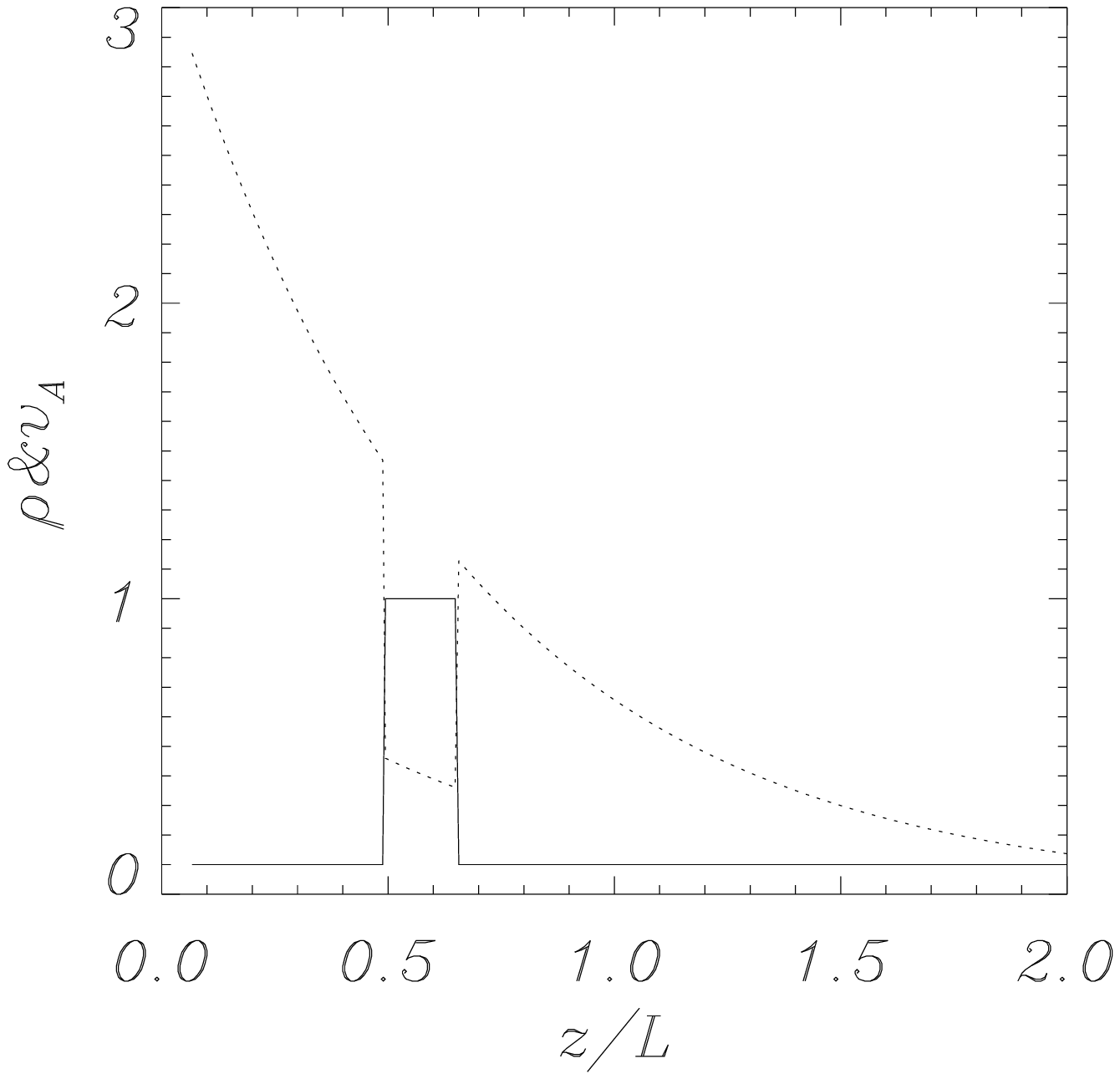}
\caption{Vertical variation along the $z$-axis of the density (solid line) and Alfv\'en speed (dotted line) for the coronal loop configuration of Figure~\ref{fig:density}.}
\label{fig:densva}
\end{center}
\end{figure}

In this paper gravity is neglected and the $\beta=0$ approximation is used. Under these assumptions the equilibrium density, $\rho$, can be chosen arbitrarily. We consider a loop with uniform density, $\rho_{0}$, embedded in a corona whose density, $\rho_{e}$, is also uniform and smaller than that of the loop by a factor 10, i.e., $\rho_{e}=\rho_{0}/10$; see Figure~\ref{fig:density}a. The vertical density profile at the arcade center is shown in Figure~\ref{fig:densva}.

The combination of the magnetic field of Equation~(\ref{eq:arccomp}) with this sharp density profile leads to the following Alfv\'en speed distribution 
\begin{equation} 
v_{A}(x,z)=\left\{\begin{array}{ll}
v_{A0}\exp{\left(-\frac{z}{\Lambda_{B}}\right)},& \textrm{inside the loop},\\
v_{Ae}\exp{\left(-\frac{z}{\Lambda_{B}}\right)},& \textrm{otherwise},\\
\end{array}\right.
\label{eq:valfven}
\end{equation}
where $v_{A0}=B_{0}/\sqrt{\rho_{0}\mu_{0}}$ and $v_{Ae}=B_{0}/\sqrt{\rho_{e}\mu_{0}}$ are the Alfv\'en speed inside and outside the loop at the base of the corona ($z=0$). This formula gives $v_A$ at any point in the $xz$-plane. Notice that the Alfv\'en speed varies both along and across magnetic field lines in our curved configuration. 

\citet{OHP1996} and \citet{AOB2004} showed that an appropriate description of normal modes in a curved structure, such as the one considered here, can be obtained by solving the MHD equations in flux coordinates, which are determined by the previous election of the equilibrium field in Equation~(\ref{eq:arccomp}). Appropriate flux coordinates are given by the following expressions

\begin{equation} 
\psi(x,z)=\cos\left(\frac{x}{\Lambda_{B}}\right)\exp\left({-\frac{z}{\Lambda_{B}}}\right),\quad\quad\quad0\leq \psi \leq1,
\label{eq:psicoord}
\end{equation}
\begin{equation}
\chi(x,z)=\frac{\sin\left(\frac{x}{\Lambda_{B}}\right)\exp\left({-\frac{z}{\Lambda_{B}}}\right)}{(1-\psi^{2})^{1/2}},\quad\quad\quad-1\leq \chi \leq1,
\label{eq:chicoord}
\end{equation}
where $\psi$ and $\chi$ are the coordinates across and along the equilibrium magnetic field lines. The normalization of the length of all field lines to the same value is achieved by the factor $(1-\psi^{2})^{1/2}$ in the denominator of Equation~(\ref{eq:chicoord}). One of the advantages of using this coordinate system is that it enables to include the whole coronal arcade in the finite domain $\psi \in [0,1]$, $\chi \in [-1,1]$. In this domain magnetic field lines are straight and each of them is represented by a different value of $\psi$. In addition, the magnetic field strength depends on both $\psi$ and $\chi$. The shape of the potential arcade and the coronal loop in these field related coordinates is shown in Figure~\ref{fig:density}b.

\section{Magnetohydrodynamic equations and linear waves}
\label{eq_linear}
In order to study small-amplitude oscillations in our potential arcade with an embedded loop the previous equilibrium is perturbed. For linear and adiabatic MHD perturbations in the zero-$\beta$ approximation the relevant equations are
 
\begin{equation} 
\rho\frac{\partial\mathbf{v}_{1}}{\partial t}=\frac{1}{\mu_{0}}(\nabla\times\mathbf{B}_{1})\times\mathbf{B},
\label{eq:momentum}
\end{equation}

\begin{equation} 
\frac{\partial \mathbf{B}_{1}}{\partial t} =\nabla\times(\mathbf{v}_{1}\times\mathbf{B}),
\label{eq:induction}
\end{equation}
where $\rho$ and $\mathbf{B}$ are the equilibrium density and magnetic field and the subscript 1 is used to represent the perturbed velocity, $\mathbf{v}_{1}$, and magnetic field, $\mathbf{B}_{1}$. 

When these equations are particularized to our two-dimensional equilibrium it turns out that it is advantageous to use field-related components instead of Cartesian components in order to characterize the directions of interest related to the polarization of each wave type. The unit vectors in the directions normal, perpendicular, and parallel to the equilibrium magnetic field are given by
\begin{equation}
\begin{array}{ll}
\hat{e}_{n}&=\frac{\nabla A}{ \vert \nabla A\vert}, \\
\\
\hat{e}_{\bot}&=\hat{e}_{y}, \\
\\
\hat{e}_{\Vert}&=\frac{\mathbf{B}}{\vert \mathbf{B}\vert},\\
\end{array}
\label{eq:unitvectors}
\end{equation}
where $A$ is the flux function, given by Equation~(\ref{eq:flux}), and $\vert \mathbf{B}\vert=B_{0}\exp\left(-z/\Lambda_{B}\right)$ is the magnetic field strength. In a low-$\beta$ plasma and in the absence of perpendicular propagation, these three directions are associated with the velocity perturbation of the three types of waves that can be excited, namely $v_{1n}$ for fast waves, $v_{1\bot}$ for Alfv\'en waves, and $v_{1\Vert}$ for slow waves.

As the equilibrium is invariant in the $y$-direction, we can Fourier analyze all perturbed quantities in the $y$-direction by making them proportional to $\exp{(ik_{y}y)}$, where $k_{y}$ is the perpendicular wave number. In this way, three-dimensional propagation is allowed and each Fourier component can be studied separately. By making the appropriate definitions, the field-related components of the MHD wave equations can be cast in the following manner

\begin{eqnarray}
\frac{\partial v_{1n}}{\partial t}&=&\frac{\vert \mathbf{B}\vert}{\mu_{0}\rho}\Bigg[\left(B_{a}-\psi_{a}\partial_{n}-\chi_{a}\partial_{\Vert}\right)B_{1\Vert}+\left(\psi_{s}\partial_{n}+\chi_{s}\partial_{\Vert}-B_{s}\right)B_{1n}\Bigg],\label{eq:velocityn}\\
\frac{\partial v_{1y}}{\partial t}&=&\frac{\vert \mathbf{B}\vert}{\mu_{0}\rho}\Bigg[\left(\psi_{s}\partial_{n}+\chi_{s}\partial_{\Vert}\right)B_{1y}+k_{y}B_{1\Vert}\Bigg],\label{eq:velocityy}\\
\frac{\partial B_{1n}}{\partial t}&=&\vert \mathbf{B}\vert\left(\psi_{s}\partial_{n}+\chi_{s}\partial_{\Vert}+B_{s}\right)v_{1n},\label{eq:fieldn}\\
\frac{\partial B_{1y}}{\partial t}&=&\vert \mathbf{B}\vert\left(\psi_{s}\partial_{n}+\chi_{s}\partial_{\Vert}\right)v_{1y},\label{eq:fieldy}\\
\frac{\partial B_{1\Vert}}{\partial t}&=&-\vert \mathbf{B}\vert\Bigg[k_{y}v_{1y}+\left(\psi_{a}\partial_{n}+\chi_{a}\partial_{\Vert}+B_{a} \right)v_{1n}\Bigg]. \label{eq:fieldpar}\\
\nonumber
\end{eqnarray}

The notation used to designate the different components of the fields and coordinates is related to the directions defined by the equilibrium field, so that the field components with subscript $n$, $y$, and $\Vert$ refer to the directions normal, perpendicular, and parallel to the equilibrium field lines. To unify the notation, the partial derivatives related to the flux coordinates $\psi$ and $\chi$ have been also rewritten with the subscript $n$ and $\Vert$ respectively, i.e.

\begin{equation}
\begin{array}{ll}
\partial_{n}&\equiv \hat{e}_{n}\cdot \nabla, \\
\\
\partial_{y}&\equiv ik_{y}\equiv \hat{e}_{y}\cdot \nabla, \\
\\
\partial_{\Vert}&\equiv \hat{e}_{\Vert}\cdot \nabla.\\
\end{array}
\label{eq:derivunitvectors}
\end{equation}

Furthermore, the quantities $\psi_{s}$, $\psi_{a}$, $\chi_{s}$, $\chi_{a}$, $B_{s}$, and $B_{a}$, that represent the derivatives of the flux coordinates and the magnetic field strength along (subscript 's') and across (subcript 'a') magnetic field lines, are defined in \citet{OHP1996}, 

\begin{equation}
\begin{array}{ll}
\psi_{a}=\frac{1}{B}(\nabla A\cdot \nabla)\psi, &\psi_{s}=\frac{1}{B}(\mathbf{B}\cdot \nabla)\psi,\\
\\
\chi_{a}=\frac{1}{B}(\nabla A\cdot \nabla)\chi, &\chi_{s}=\frac{1}{B}(\mathbf{B}\cdot \nabla)\chi,\\
\\
B_{a}=\frac{1}{B^{2}}(\nabla A\cdot \nabla)B, &B_{s}=\frac{1}{B^{2}}(\mathbf{B}\cdot \nabla)B.\\
\end{array}
\label{eq:defpsischis}
\end{equation}

Notice that such a general derivation and implementation of the governing equations enables us to apply the scheme to any other configuration for which similar flux coordinates can be defined.

Equations~(\ref{eq:velocityn})--(\ref{eq:fieldpar}) constitute a set of coupled partial differential equations with non-constant coefficients that describe the propagation of fast and Alfv\'en waves. As the plasma $\beta$ is zero, slow waves are excluded from the analysis and $v_{1\Vert}=0$. When $k_{y}=0$, Equations~(\ref{eq:velocityn})--(\ref{eq:fieldpar}) constitute two independent sets of equations. The two equations for $v_{1y}$ and $B_{1y}$ are associated to Alfv\'en wave propagation. On the other hand, the three equations for the remaining variables, $v_{1n}$, $B_{1n}$, $B_{1\Vert}$, describe fast wave propagation.

\section{Numerical method and initial and boundary conditions}
\label{method}
The obtained set of differential equations are too complicated to have analytical or simple numerical solutions. For this reason we have solved them with a numerical code described in \citet{BBT2009}, using flux coordinates. This particular choice of coordinates is crucial for the proper computation of the solutions, and has been overlooked in previous numerical studies.

We try to mimic the observed vertical oscillations of coronal loops in the corona when a sudden release of energy occurs and perturbs the structure. Since we are concerned with the excitation of fast waves, our initial perturbation is such that only the normal velocity component is disturbed, whereas all the other variables ($v_{1y}$, $B_{1n}$, $B_{1y}$, $B_{1\Vert}$) are initially set to zero. To initially excite the system we have chosen a two-dimensional profile in $v_{1n}$ with, respectively, a cosine function and a Gaussian profile along and across field lines,

\begin{equation}
v_{1n}(\chi,\psi)=v_{0}\cos(k_{\Vert}\chi)\exp\left[-\left(\frac{\psi-\psi_{0}}{a}\right)^{2}\right].
\label{eq:initcond}
\end{equation}
This expression represents an initial disturbance that perturbs a range of field lines centered about the field line $\psi=\psi_{0}$. Here $v_{0}$ is the initial amplitude of the disturbance, $a$ is related to the range of field lines initially affected by the perturbation, and $k_{\Vert}$ is the wave number of the disturbance along the magnetic field. This symmetric initial profile does not represent a general disturbance in the solar corona, but it is selected in order to mainly excite the fundamental fast mode with one maximum along field lines. This can be achieved by an adequate choice of $k_{\Vert}$. A more general disturbance would excite an ensemble of modes at the same time, making the interpretation of the obtained numerical results more difficult.

The implementation of the appropriate boundary conditions in the numerical code is an important issue. The reflection of waves at the bottom boundary, due to the large inertia of the photospheric plasma, is accomplished by imposing line-tying boundary conditions at $z=0$. In all other boundaries, flow-through conditions are imposed so that perturbations are free to leave the system. As the equations are solved in flux coordinates, one then needs to make the proper translation between the system boundaries from Cartesian to field related coordinates; see Figure~\ref{fig:density}. Although the whole arcade can be reproduced in flux coordinates, considering the complete range $0\leq\psi\leq 1$ causes numerical issues, so we restrict ourselves to the range $[0.034,0.9]$ in the $\psi$-direction. This implies that extremely low and extremely high field lines are discarded in the numerical simulations.

\section{Numerical results}
\label{results}
\begin{figure}[h]
\begin{center}
\includegraphics[width=0.5\textwidth]{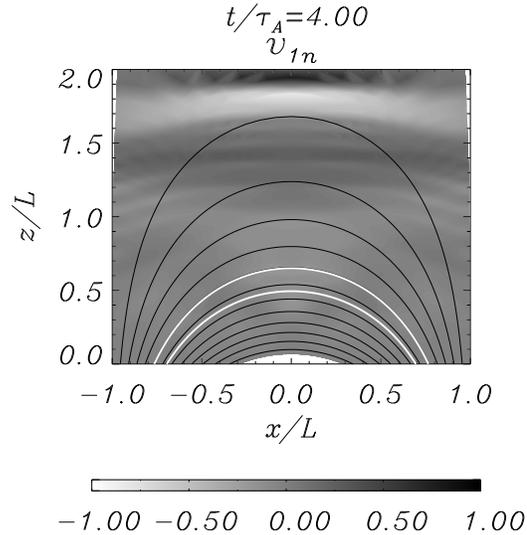}
\caption{Snapshot of the two-dimensional distribution of the normal velocity component for $k_{y}=0$. Some magnetic field lines (black curves) and the edge of the coronal loop (white lines) have been represented. (An animation of this figure is available in the online journal.)}
\label{fig:movies1}
\end{center}
\end{figure}

The results presented in this section have been obtained with the numerical solution of Equations~(\ref{eq:velocityn})--(\ref{eq:fieldpar}) after an initial perturbation given by Equation~(\ref{eq:initcond}) with $\psi_{0}=0.41$ and $a=0.05$ is launched. Different values of $k_{y}$ have been considered ($k_yL=0,5,16,60$). The two-dimensional variation of $v_{1n}$ and $v_{1y}$ for some of these simulations is presented as animations associated to Figures~\ref{fig:movies1} and \ref{fig:movies2}. Time in these animations and in subsequent plots is given in units of $\tau_A=L/v_{A0}$.

\subsection{Wave propagation in the plane of the arcade --- wave leakage}
\label{sect:wave leakage}
\begin{figure}[h]
\begin{center}
\includegraphics[width=0.49\textwidth]{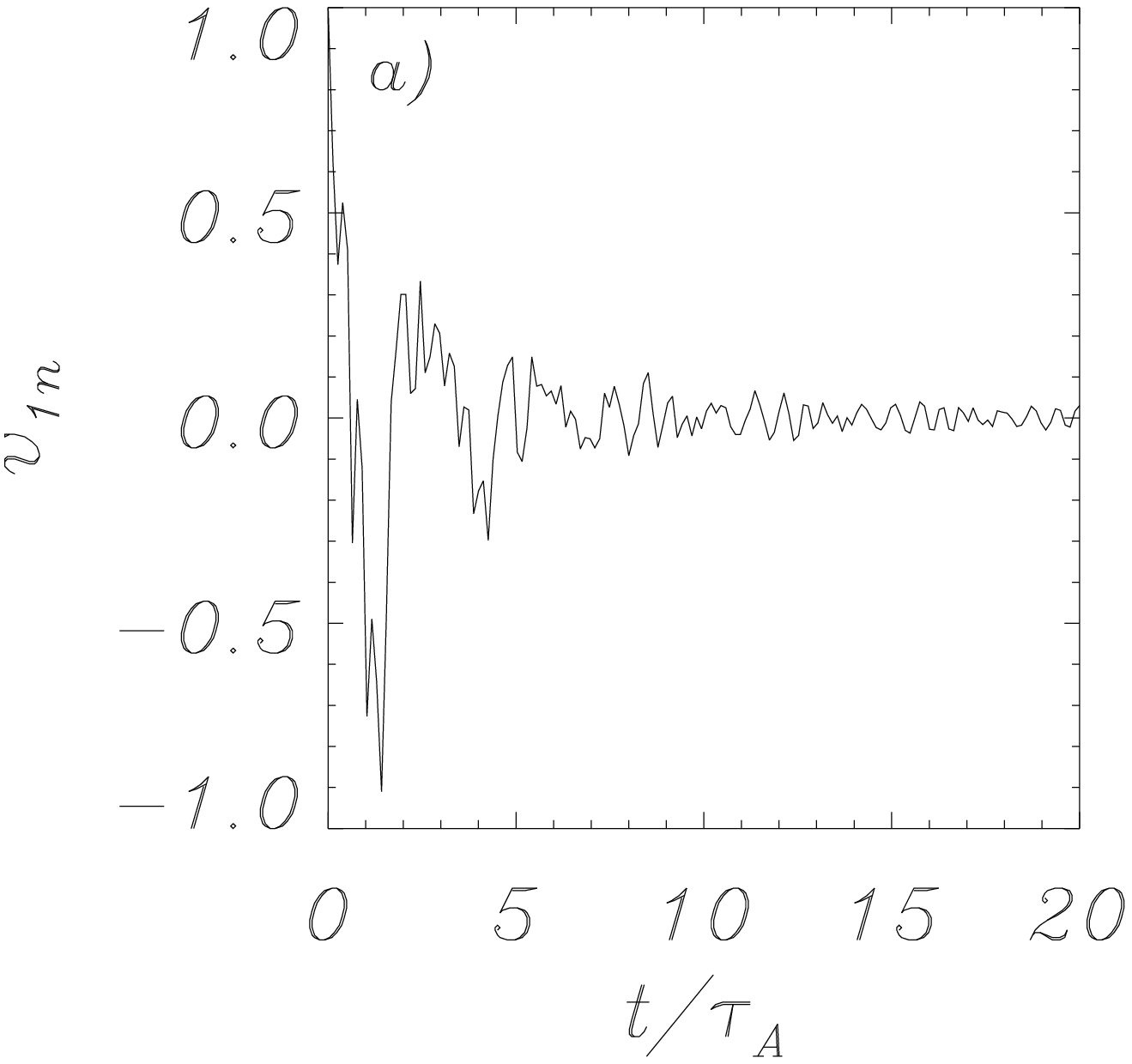}
\includegraphics[width=0.49\textwidth]{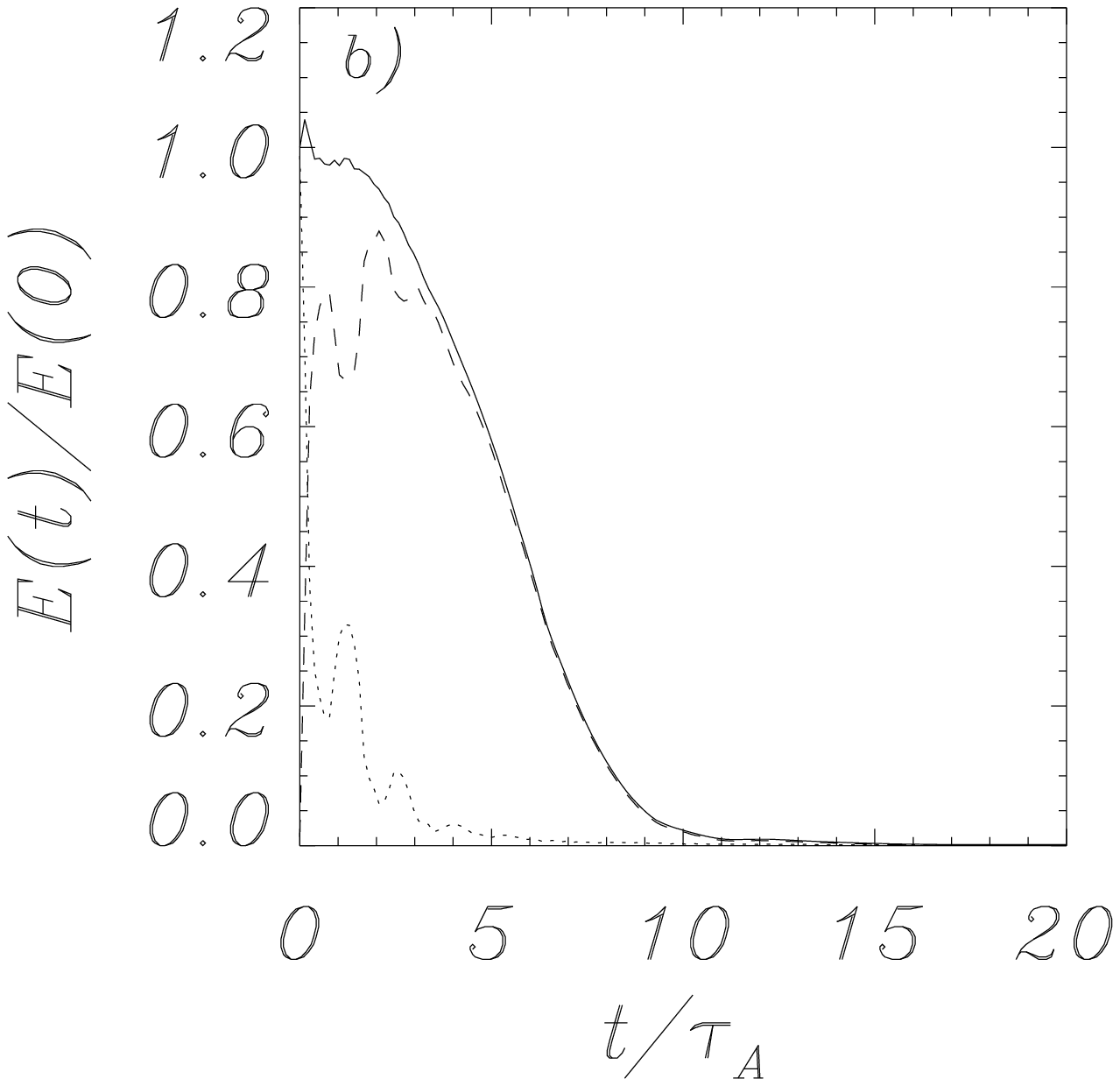}
\caption{Results for $k_y=0$. (a) Temporal evolution of the normal velocity component, $v_{1n}$, at the loop center ($x=0$, $z/L=0.56$). (b) Normalized energy density in the whole domain (solid thick line), inside the loop (dotted line), and outside the loop (dashed line) as a function of time. The two-dimensional temporal evolution of $v_{1n}$ is presented in the animation associated to Figure~\ref{fig:movies1}.}
\label{fig:ky0}
\end{center}
\end{figure}

We first consider wave propagation in the plane of the arcade ($k_{y}=0$). Such as can be appreciated in the animation accompanying Figure~\ref{fig:movies1}, the initial perturbation produces traveling disturbances across the magnetic surfaces that propagate away from the density enhancement. After a short time ($t/\tau_A\sim 5$) the loop contains very little energy, and so these disturbances can be interpreted as a combination of leaky modes that the loop structure is unable to confine. This interpretation is also supported by the time evolution of the normal velocity component at the center of the loop (Figure~\ref{fig:ky0}a), that displays a strong damping such that, for $t/\tau_A\gtrsim 5$, the velocity has an almost null amplitude inside the loop.

A convenient way to quantify the wave energy leakage and the damping is to use the total energy density computed both in the full domain and in the interior of the loop. We use the following formula to calculate the energy density at a given position and time,

\begin{equation}
\delta E(\mathbf{r},t)=\frac{1}{2}\left[\rho\left(v_{1n}^2+v_{1y}^2\right)+\frac{1}{\mu_{0}}\left(B_{1n}^2+B_{1y}^2+B_{1\Vert}^2\right)\right].
\label{eq:totalenergy}
\end{equation}
The total energy density in a spatial domain, $D$, of the $xz$-plane can be computed from the spatial integration of $\delta E(\mathbf{r},t)$,
\begin{equation}
E(t)=\int_D\delta E(\mathbf{r},t)\,dx\,dz.
\label{eq:totalenergydomain}
\end{equation}
In our plots this quantity is normalized by dividing it by the initial energy density over the whole numerical domain, $E(0)$.

Figure~\ref{fig:ky0}b shows the total energy density integrated over the full domain, the interior of the loop, and the external region. We can see that the loop is unable to trap wave energy and that, after a little more than $5\tau_A$, it has already transferred all its energy to the outside medium. After this time, the wave energy outside the loop equals the energy in the full domain. Because of this wave leakage, the total energy of the system continuously decreases and when the last leaky waves carrying a non-negligible amount of energy reach the domain boundaries (i.e. at $t/\tau_A\sim 10$), it is practically zero; see also the animation of Figure~\ref{fig:movies1}.

These results differ from what is obtained in a line-tied straight slab model of a coronal loop \citep{TOB2005a}, in which the kink mode of the loop remains oscillating after the initial leaky phase, since the density enhancement is able to trap part of the energy of the initial perturbation. Such as was already pointed out by \citet{MSN2005,SMS2005,SSM2006,VFN2006a,VFN2006b,VFN2006c,SMS2007}, the magnetic field line curvature produces damping by wave leakage and the energy deposited initially in the loop is simply radiated as a combination of leaky modes, and thus no trapped solutions are found.

\subsection{Wave propagation in three-dimensions --- wave trapping}
\label{sect:wave trapping}

\citet{TOB2006a} showed that in a toroidal loop structure wave leakage is much less effective, that in slab geometry. Following the same idea, \citet{ATO2007} used a simple three-dimensional straight slab model to demonstrate that in the case of a slab, introducing oblique propagation increases the spatial confinement of the kink mode around the loop. \citet{VVT2009} have also shown that a cylindrical tube is a more efficient wave guide than the slab model, and less energy is allowed to leak. When a potential arcade is used as the equilibrium structure, the oblique propagation of the slab corresponds to perpendicular propagation, which means the introduction of the $k_{y}$ wave number. If the previous results for the kink mode with oblique propagation in a coronal slab are also valid in the present loop configuration, we expect that an initial perturbation will excite both leaky waves (that will leave the system in a similar manner as found in \S~\ref{sect:wave leakage}) and trapped modes (that will retain energy in the system). A comment about the term ``trapped'' mode used in this paper is in order. In our plasma configuration the local cut-off frequency depends on position because of the Alfv\'en speed inhomogeneity and the inclusion of perpendicular wave propagation. For this reason, an eigenfunction that is evanescent in a certain region can change its character to propagating in another part of the system, and so wave energy leakage can arise. For this reason we warn that modes termed ``trapped'' here may actually loose energy by this mechanism. If leakage is very small the modes have, to all practical purposes, a trapped behaviour.

\begin{figure}[h]
\begin{center}
\includegraphics[width=0.7\textwidth]{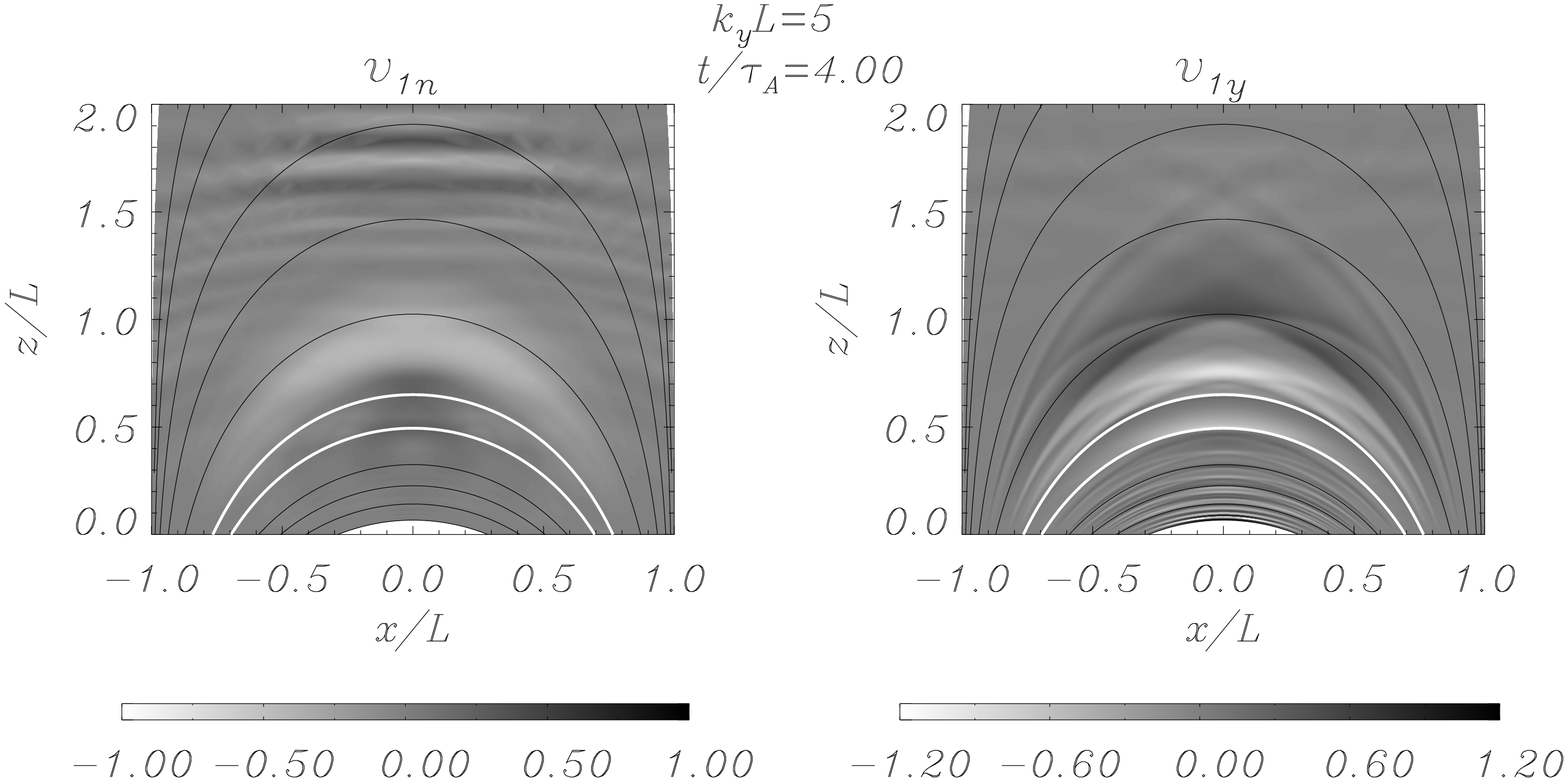}\\
\includegraphics[width=0.7\textwidth]{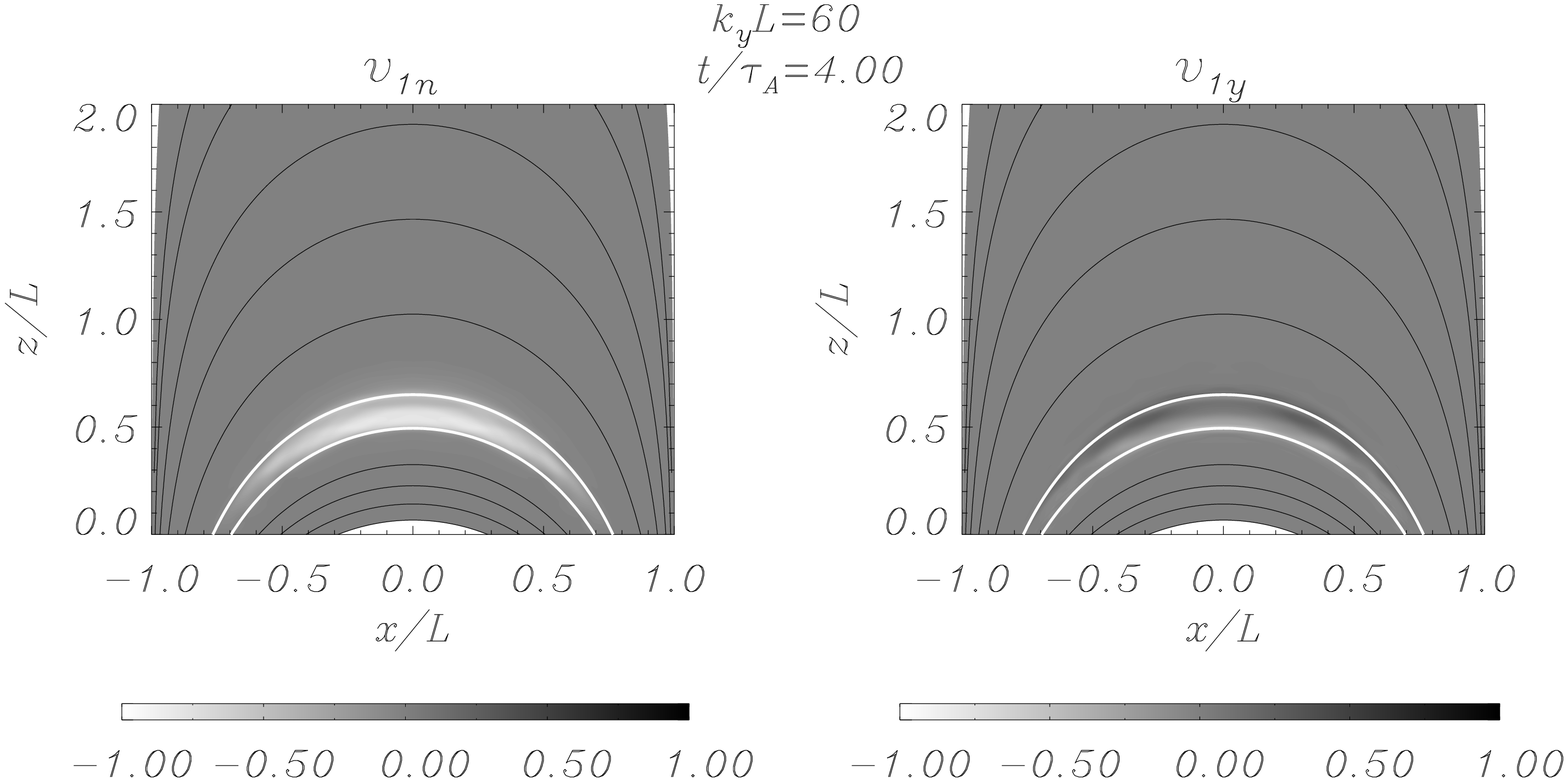}
\caption{Snapshots of the two-dimensional distribution of the normal and perpendicular velocity components for $k_{y}L=5$ and $k_{y}L=60$. Some magnetic field lines (black curves) and the edge of the coronal loop (white lines) have been represented. (An animation of this figure is available in the online journal.) }
\label{fig:movies2}
\end{center}
\end{figure}

As in \S~\ref{sect:wave leakage}, we have considered an impulsive excitation of the normal velocity component, but now $k_yL=5$. The temporal evolution of the normal and perpendicular velocity components (see the animations accompanying the top panel of Figure~\ref{fig:movies2}) illustrate that there are important differences compared to the case $k_y=0$ (Figure~\ref{fig:movies1}). First of all, some of the energy contained in the initial perturbation goes to the perpendicular variables because $k_y\neq 0$ implies that the perturbed perpendicular components ($v_{1y}$ and $B_{1y}$) are no longer independent of the normal and parallel ones ($v_{1n}$, $B_{1n}$, and $B_{1\Vert}$). In addition, we have just discussed that for $k_y=0$ most energy has left the system at $t/\tau_A\sim 10$. For $k_yL=5$, however, it is apparent that some energy remains in the loop and part of the arcade above it for much longer times. This is interpreted as a signature of the excitation of one or more trapped modes of the system. We next substantiate this claim with a careful inspection of $v_{1n}$.

\begin{figure}[h]
\begin{center}
\includegraphics[width=0.3\textwidth]{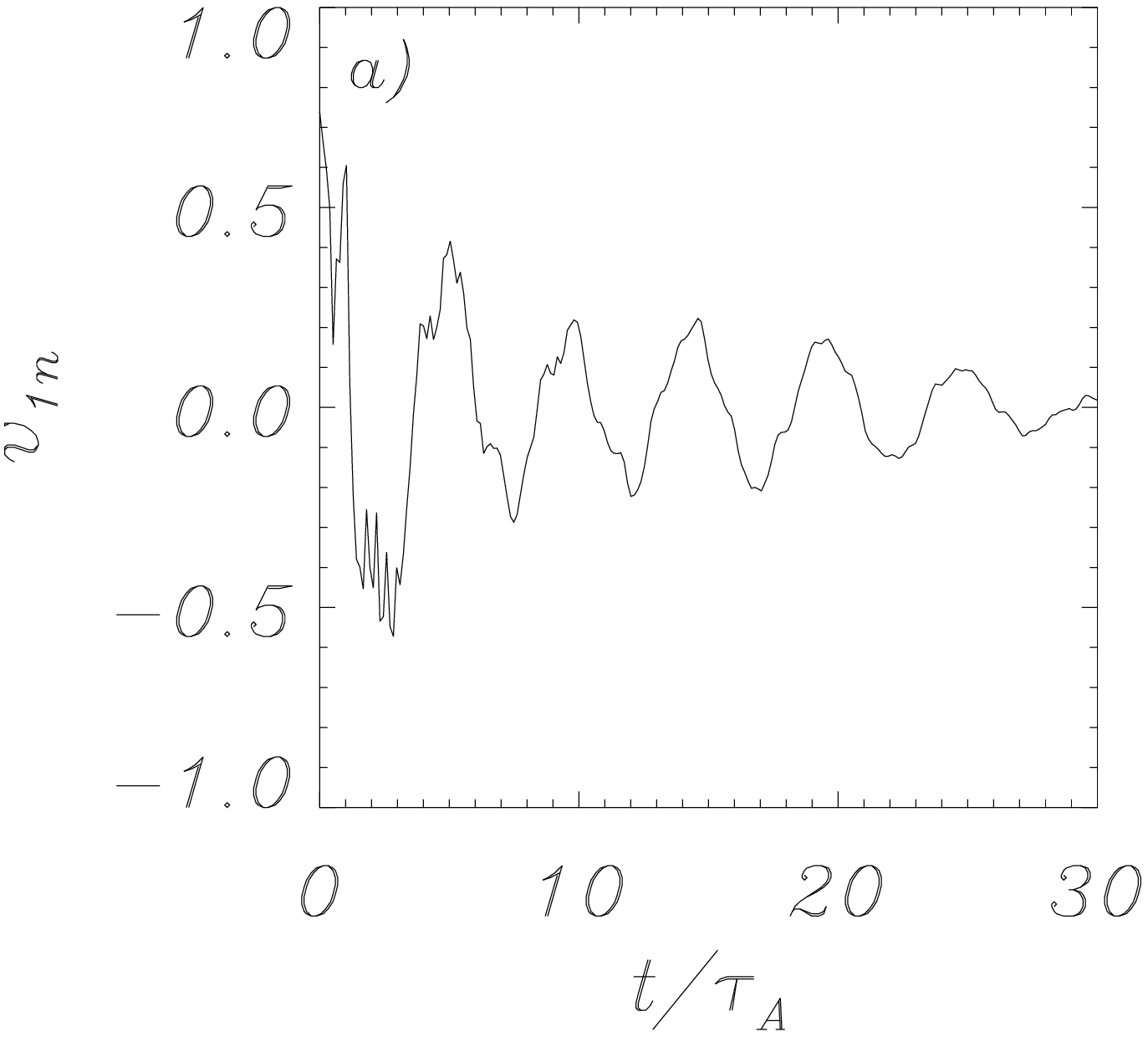}
\includegraphics[width=0.3\textwidth]{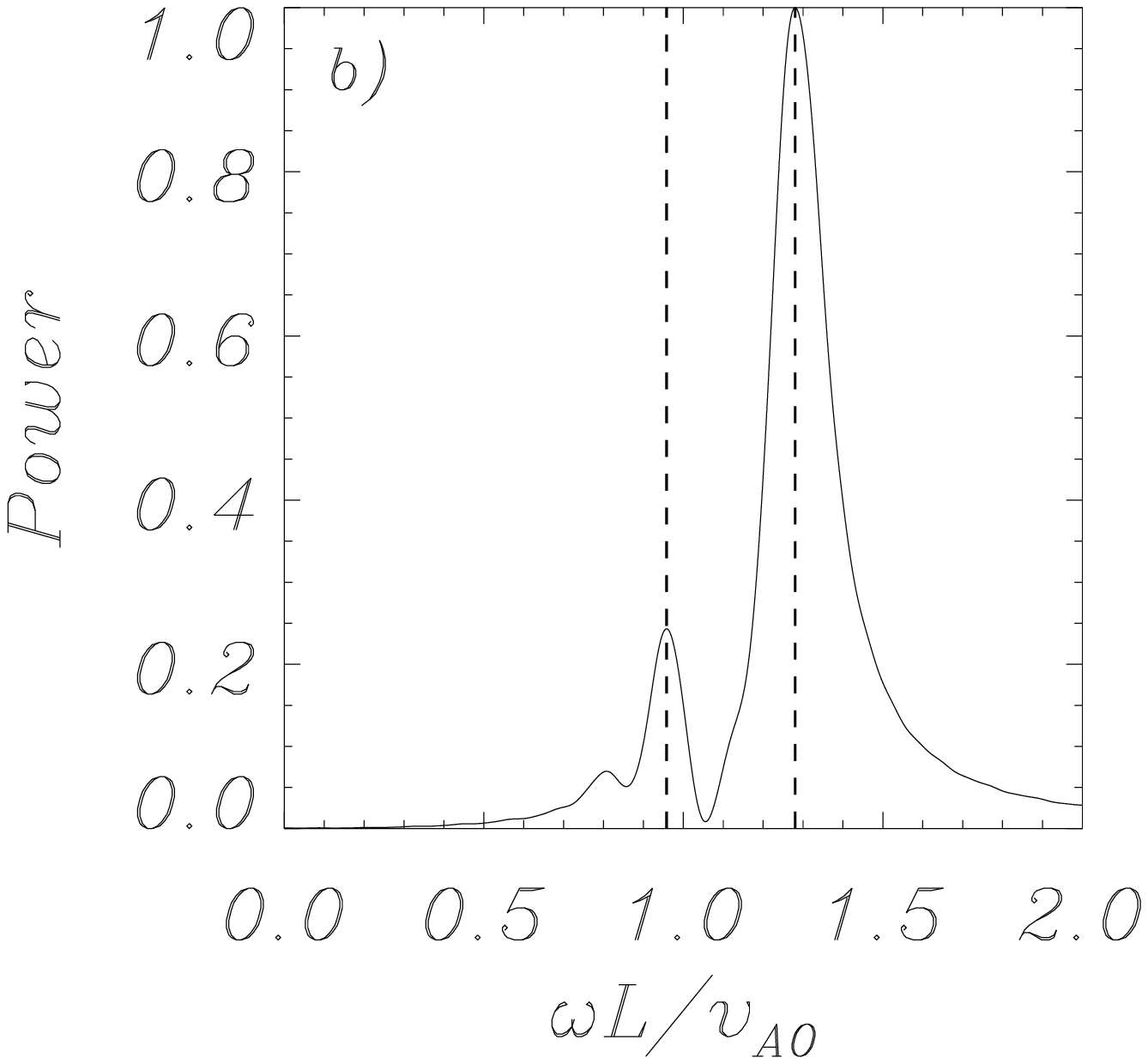}
\includegraphics[width=0.3\textwidth]{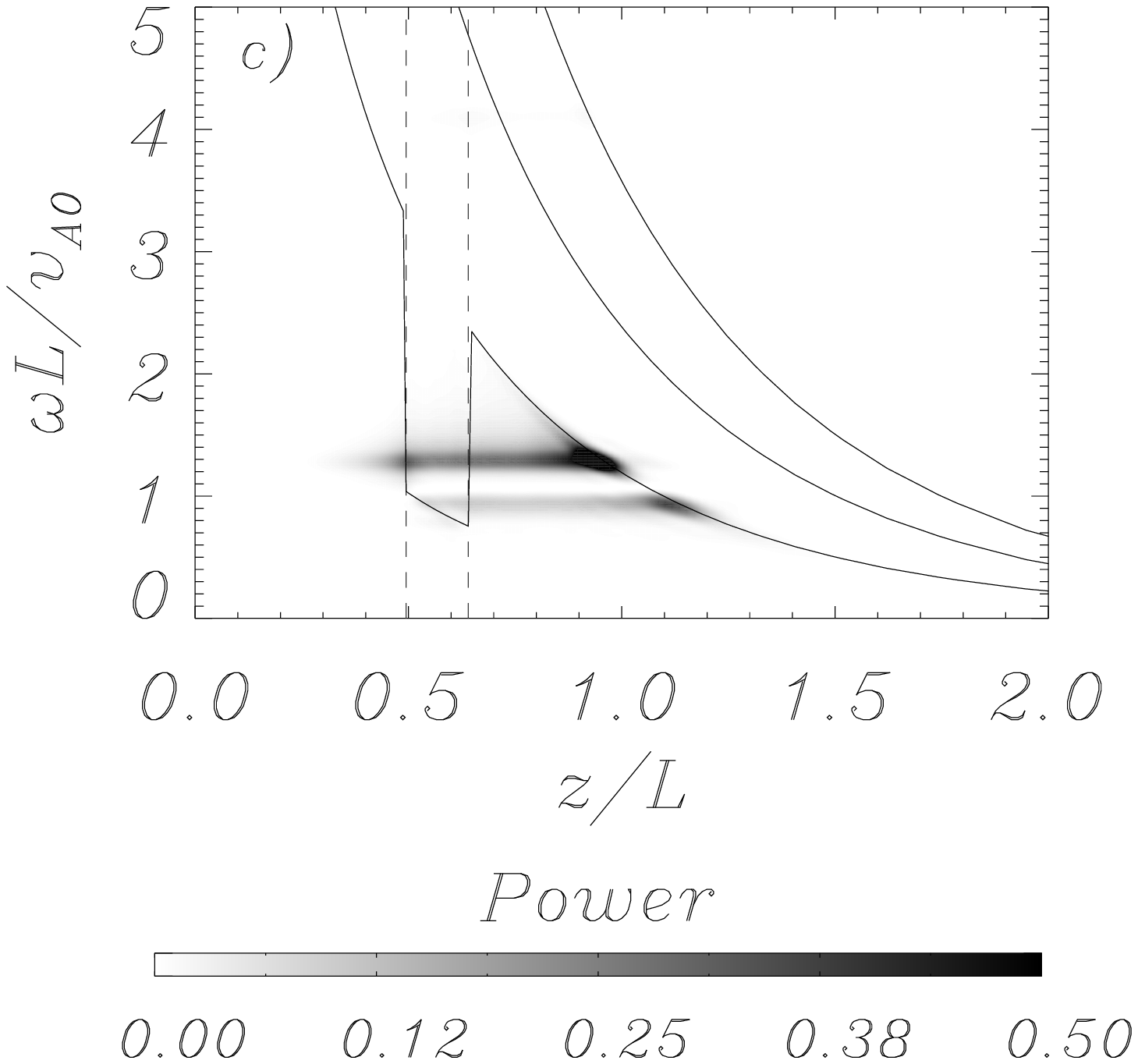}
\caption{Results for $k_{y}L=5$. (a) Temporal evolution of $v_{1n}$ near the loop center ($x=0$, $z/L=0.52$). (b) Normalized power spectrum of the signal of panel (a). (c) Power spectrum of $v_{1n}$ at $x=0$ as a function of height. The three solid lines are, from bottom to top, the frequency of the fundamental Alfv\'en mode and its first two harmonics. The two dashed lines mark the limits of the coronal loop.}
\label{fig:vnky5}
\end{center}
\end{figure}

We consider the temporal variation of the normal velocity component near the loop center (Figure~\ref{fig:vnky5}a). It displays two distinct behaviors: for $t/\tau_A\lesssim 5$ very short periods are prominent. They correspond to leaky modes, that propagate quickly away from the loop. After this leaky phase all that remains is a gentle, damped oscillation. This damping will be later discussed in detail, but let us mention that it is milder than that of the $k_y=0$ case (Figure~\ref{fig:ky0}a). The power spectrum of the $v_{1n}$ signal (Figure~\ref{fig:vnky5}b) shows two clear peaks that, according to the interpretation in the previous paragraph, simply reflect the power contained in two trapped modes excited by the initial perturbation. An evidence of this would be the presence of power peaks at nearly the same frequency in neighboring points. To investigate this possibility, we consider $v_{1n}$ as a function of $t$ for $x=0$ and different values of $z$, compute their power spectra and stack them in a contour plot (Figure~\ref{fig:vnky5}c). The result is the presence of significant power at two horizontal contours whose frequencies coincide with those found in Figure~\ref{fig:vnky5}b, each of them with its characteristic range of heights. From Figure~\ref{fig:vnky5}c we see that the fundamental mode frequency is $\omega L/v_{A0}\sim 0.95$ and that this mode covers the range $0.5\lesssim z/L\lesssim 1.15$ for $x=0$. The first harmonic has $\omega L/v_{A0}\sim 1.28$ and covers the range $0.3\lesssim z/L\lesssim 0.95$ for $x=0$.

Hence, we conclude that two trapped modes of the system have been excited by the initial perturbation and that this is the reason why the normal velocity component is less strongly attenuated for $k_y=5$ than for $k_y=0$. Both modes are spatially distributed over the coronal loop and a large region above it, that is, perpendicular wave propagation increases the wave confinement in the present equilibrium configuration, although for $k_yL=5$ wave energy around the loop is also found.

\subsection{Mode coupling and resonant energy transfer}
\label{sect:wave coupling}

\begin{figure}[h]
\begin{center}
\includegraphics[width=0.3\textwidth]{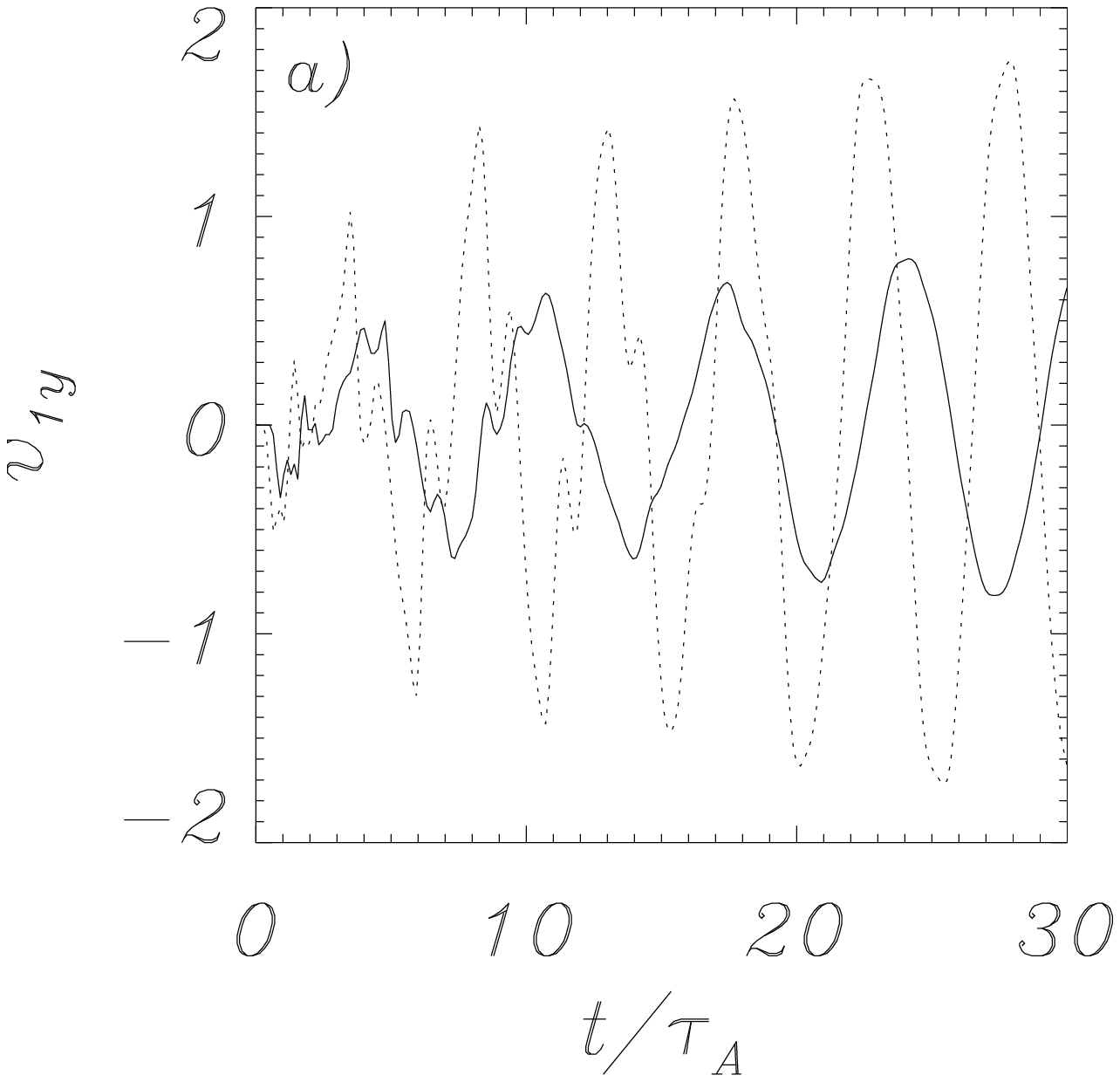}
\includegraphics[width=0.3\textwidth]{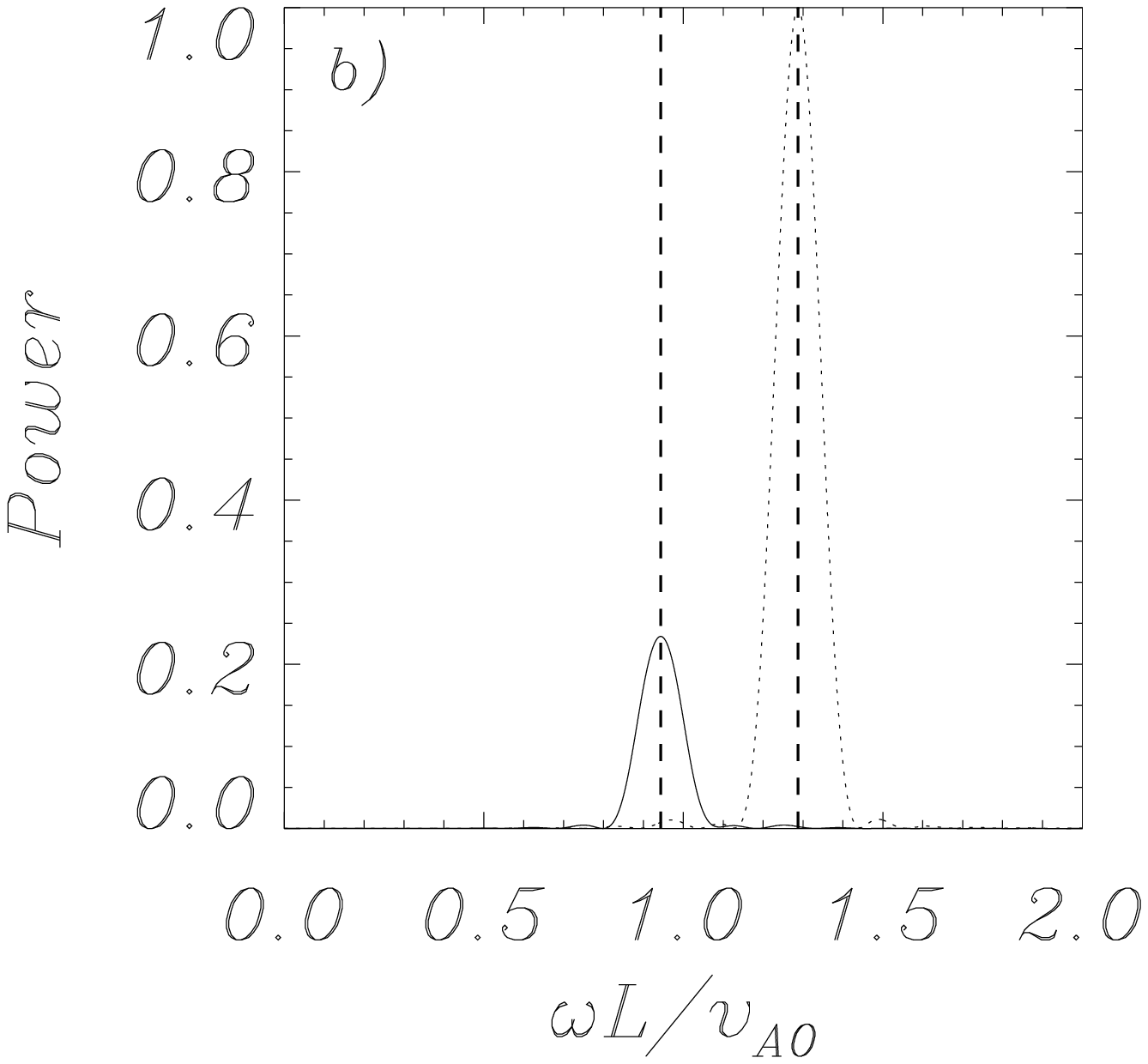}
\includegraphics[width=0.3\textwidth]{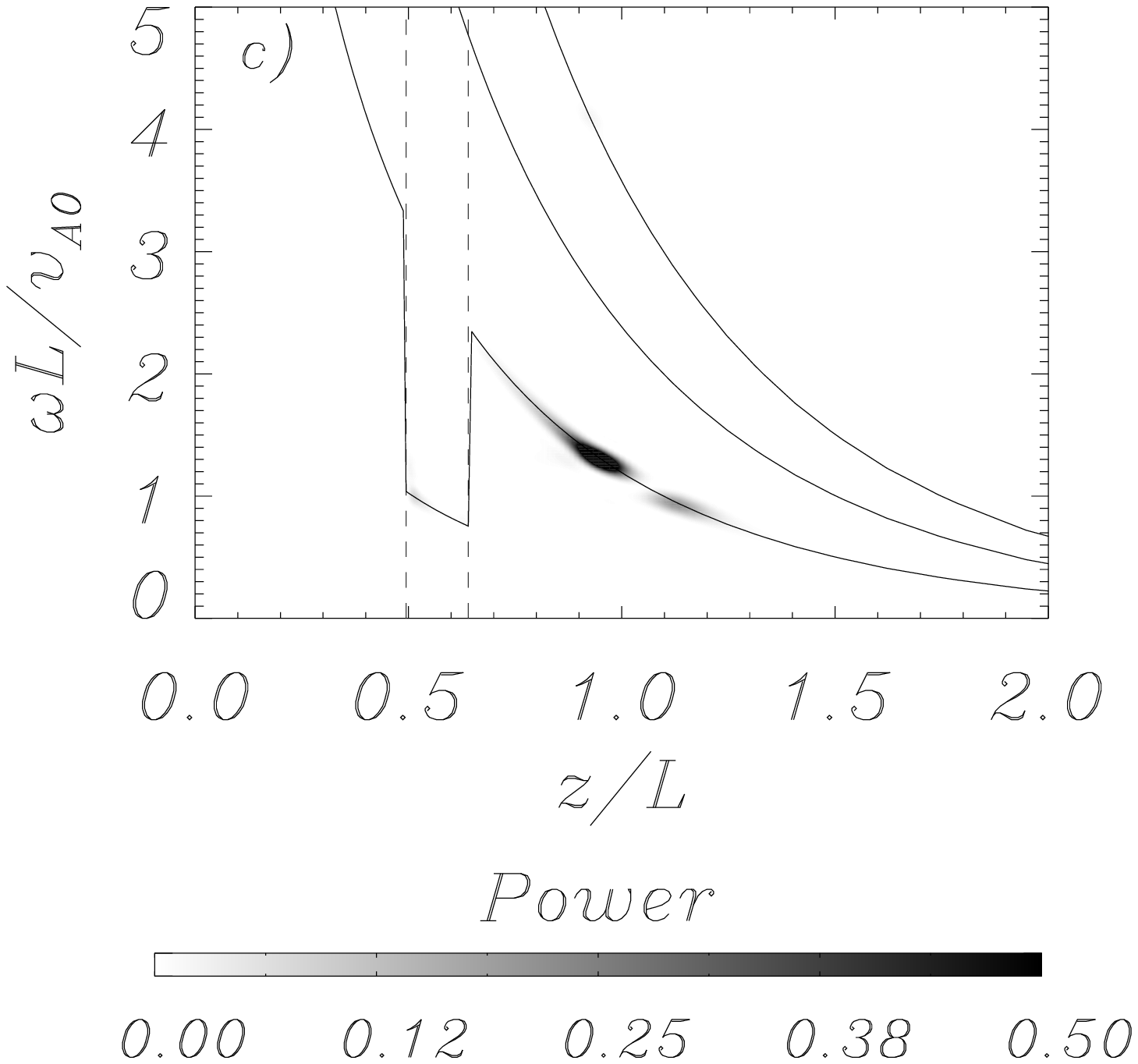}
\caption{Results for $k_yL=5$. (a) Temporal evolution of $v_{1y}$ at the points $x=0$, $z/L=1.12$ (solid line) and $x=0$, $z/L=0.96$ (dotted line). (b) Normalized power spectrum of the signals of panel (a). (c) Power spectrum of $v_{1y}$ at $x=0$ as a function of height. The three solid lines are, from bottom to top, the frequency of the fundamental Alfv\'en mode and its first two harmonics. The two dashed lines mark the limits of the coronal loop.}
\label{fig:vyky5}
\end{center}
\end{figure}

Additional information about the temporal evolution of the system comes from the perpendicular velocity component. In Figure~\ref{fig:vyky5}a this perturbed variable is plotted at two different positions at the arcade center ($x=0$). In both cases $v_{1y}$ displays an oscillatory behavior with a monotonically increasing amplitude and it is evident that the oscillatory period is different at the two positions. This is confirmed by the power spectra of these two signals, each displaying a single peak at one of the frequencies of the normal velocity components found in Figures~\ref{fig:vnky5}b and c. Next, power spectra of $v_{1y}$ as a function of $t$ along the $z$-axis are computed and stacked together (Figure~\ref{fig:vyky5}c) and it becomes clear that most of the power of this velocity component is concentrated at two particular heights. To explain this result we recall that in \S~\ref{sect:wave trapping} we mentioned that after the initial perturbation, trapped modes of the system were established by a transfer of energy from $v_{1n}$ to the other perturbed variables, including $v_{1y}$. But a second, more efficient, process takes place here: the resonant transfer of energy from the trapped modes to the Alfv\'en modes whose frequencies match those of the former. Figure~\ref{fig:vnky5}c shows that the frequency of the two prominent trapped modes found in \S~\ref{sect:wave trapping} ($\omega L/v_{A0}\sim 0.95$ and $\omega L/v_{A0}\sim 1.28$) intersect the lowest one of the three solid lines included in this plot. These solid lines are the frequencies of the Alfv\'en continua corresponding to the fundamental mode and its first two harmonics \citep{OBH1993}. Hence, we conclude that the frequencies of the two trapped modes match that of the fundamental Alfv\'en mode at two particular heights, namely $z/L\sim 1.12$ and $z/L\sim 0.96$, so that these are the positions where Alfv\'en energy appears, in agreement with Figure~\ref{fig:vyky5}c. 

It should be noted that the trapped mode frequencies also match the frequency of Alfv\'en harmonics, but this happens for heights at which the trapped mode  amplitude is negligible. \citet{AOB2004} showed that for $k_{y}\neq0$ resonant coupling between fast and Alfv\'en modes can only happen for modes with the same parity along $\bf{B}$. As a consequence, resonant absorption with Alfv\'en modes other than the fundamental one does not take place.

So far our description of the resonant absorption process is incomplete because it relies on the information along the $z$-axis. The animation of the top panel of Figure~\ref{fig:movies2} shows that the largest amplitudes of $v_{1y}$ for $t/\tau_A\gtrsim 10$ are in two ranges of magnetic surfaces centered about the field lines whose apexes are at $z/L\sim 0.96$ and $z/L\sim1.12$ (i.e. at the resonant positions in Figure~\ref{fig:vyky5}c). A relevant issue is that the Alfv\'en frequency varies in each of these two ranges of magnetic surfaces and so the resulting Alfv\'en oscillations soon become out of phase. This phase mixing creates small spatial scales that lead to the numerical dissipation of energy whose importance will next become clear.

Equation~(\ref{eq:totalenergydomain}) gives a measure of the energy density in a domain of our system from the spatial integration of Equation~(\ref{eq:totalenergy}). In these definitions, $\delta E(\mathbf{r},t)$ and $E(t)$ contain the contribution from both the Alfv\'enic perturbed variables (i.e. $v_{1y}$ and $B_{1y}$) and the perturbed variables characteristic of fast modes for $k_y=0$ (i.e. $v_{1n}$, $B_{1n}$, and $B_{1\Vert}$). We denote by $E_{Alfven}(t)$ and $E_{fast}(t)$ the spatial integral of these two contributions over the whole system,

\begin{equation}
\begin{array}{ll}
E_{Alfven}(t)=&\int\frac{1}{2}\left(\rho v_{1y}^2+\frac{1}{\mu_{0}}B_{1y}^2\right)\,dx\,dz, \hspace{1cm}\\
\\
E_{fast}(t)=&\int\frac{1}{2}\left[\rho v_{1n}^2+\frac{1}{\mu_{0}}\left(B_{1n}^2+B_{1\Vert}^2\right)\right]\,dx\,dz.
\end{array}
\label{eq:energycontrib}
\end{equation}
Since these two integrals are carried out over the full arcade, it is clear that $E_{Alfven}(t)+E_{fast}(t)$ is equal to $E(t)$ of Equation~(\ref{eq:totalenergydomain}) when $D$ is the complete numerical domain.

\begin{figure}[h]
\begin{center}
\includegraphics[width=0.45\textwidth]{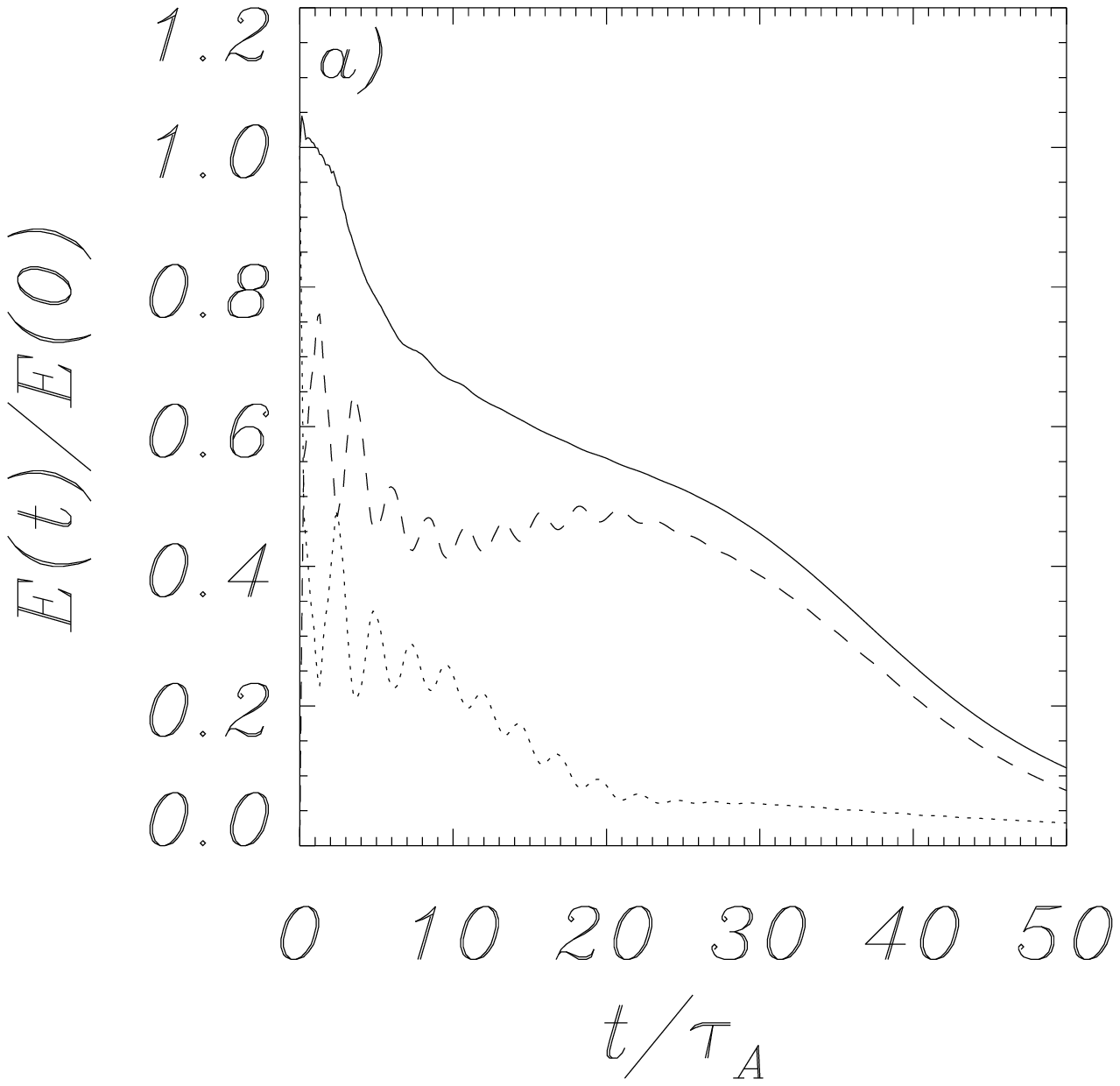}
\includegraphics[width=0.45\textwidth]{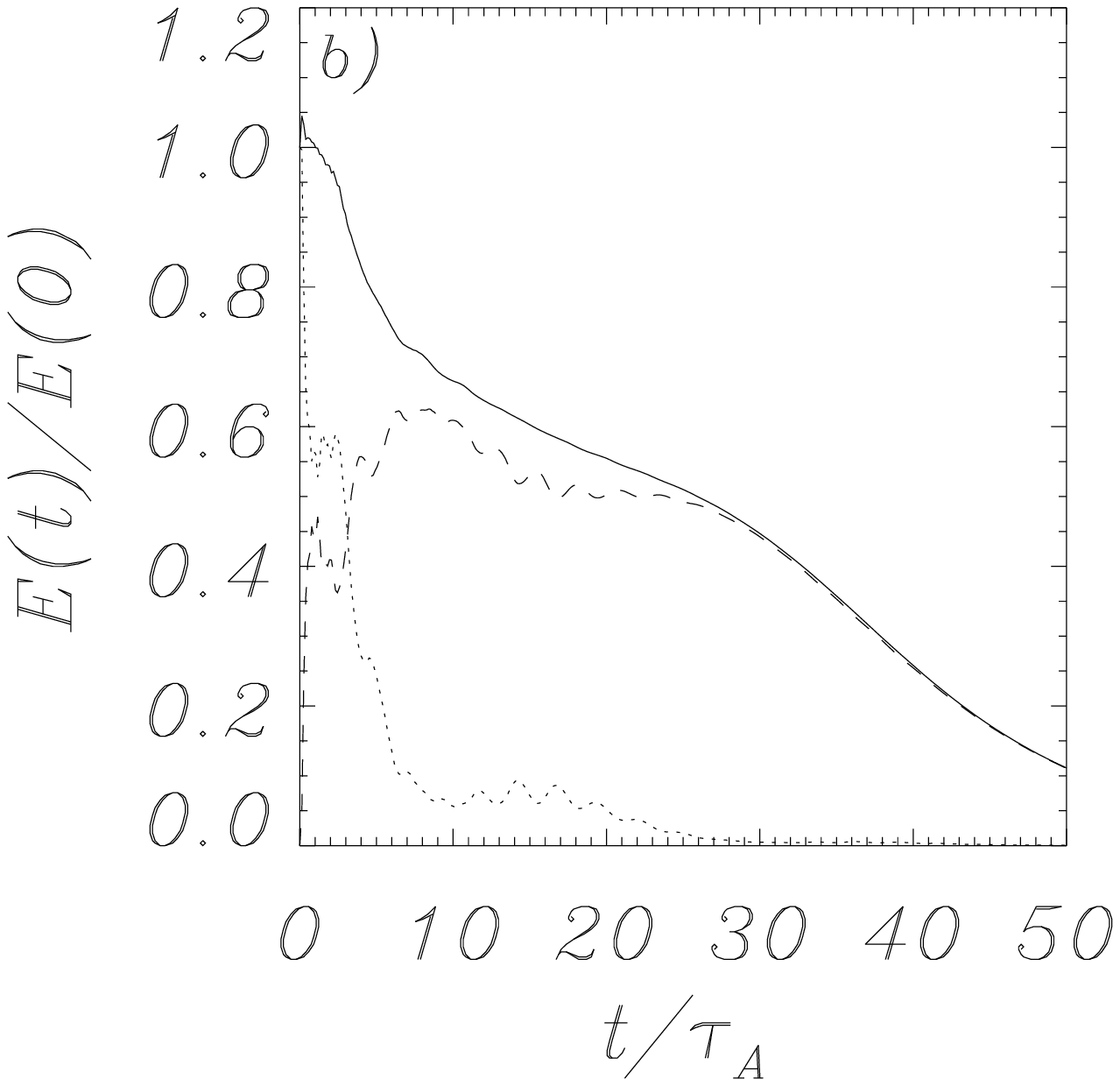}
\caption{Results for $k_yL=5$. (a) Normalized energy density integrated over the whole system (solid line), integrated over the coronal loop (dotted line), and integrated over the region outside the loop (dashed line). (b) Normalized energy density integrated over the whole system contained in all perturbed variables (solid line; same as solid line of panel (a)), in the ``fast'' perturbed variables, $E_{fast}(t)$ (dotted line), and in the Alfv\'enic perturbed variables, $E_{Alfven}(t)$ (dashed line).}
\label{fig:energiesky5}
\end{center}
\end{figure}

Figure~\ref{fig:energiesky5}a shows the energy density in the system, inside the loop and outside it. The first of these three quantities presents three phases: for $t/\tau_A\lesssim 5$ the total energy in the whole system decreases rapidly because of  the emission of leaky waves. Then, for $5\lesssim t/\tau_A\lesssim 25$ the rate of energy decrease is slower, and for $t/\tau_A\gtrsim 25$ it becomes stronger again. To understand the last two phases one must bear in mind the energy loss by the numerical dissipation of Alfv\'en waves. During the phase $5\lesssim t/\tau_A\lesssim 25$ there is a substantial amount of energy in the loop (dashed line in Figure~\ref{fig:energiesky5}a) that is slowly transferred to the surrounding medium (dotted line in Figure~\ref{fig:energiesky5}a). We know that this process is caused by resonant absorption, by which the energy that could not leak in the initial phase is transferred to the two resonant positions. After $t/\tau_A\sim 25$ almost all the energy in the trapped modes has been given to Alfv\'en modes, which implies that these modes cannot longer ``feed'' from their previous energy ``reservoir''. As a result, the numerical dissipation of phase-mixed Alfv\'en waves proceeds at a faster pace. Figure~\ref{fig:energiesky5}b provides support to this interpretation: at $t/\tau_A\sim 10$ almost all the energy has been converted from the ``fast''-like perturbations to the Alfv\'enic ones, but $E_{fast}$ does not become negligible until $t/\tau_A\sim 30$. This means that some conversion to $E_{Alfven}$ still takes place for $10\lesssim t/\tau_A\lesssim 30$, and so the dissipation of Alfv\'en waves is slower than in the later phase.

\subsection{Wave confinement and resonant absorption for large $k_y$}
\label{sect:largeky}

To evaluate the influence of the perpendicular wave number on the confinement of wave energy by trapped modes of the loop and the loss of this energy by resonant absorption to Alfv\'en continuum modes we consider $k_yL=60$. We first inspect the two-dimensional temporal evolution of the perturbed velocity components (animations associated to the bottom panel of Figure~\ref{fig:movies2}). There are obvious differences with the numerical simulation with $k_yL=5$: first of all, there is a lack of wave leakage at the beginning of the simulation, shortly after the initial impulse is released. Second, large velocity amplitudes are only found inside the loop structure, both for the $v_{1n}$ and the $v_{1y}$ components. And third, since the excited trapped modes are not spread outside the loop, there is no resonant absorption at large heights. In summary, these animations seem to indicate a much stronger confinement of the wave energy by the loop. Next, let us confirm this preliminary result.

\begin{figure}[h]
\begin{center}
\includegraphics[width=0.49\textwidth]{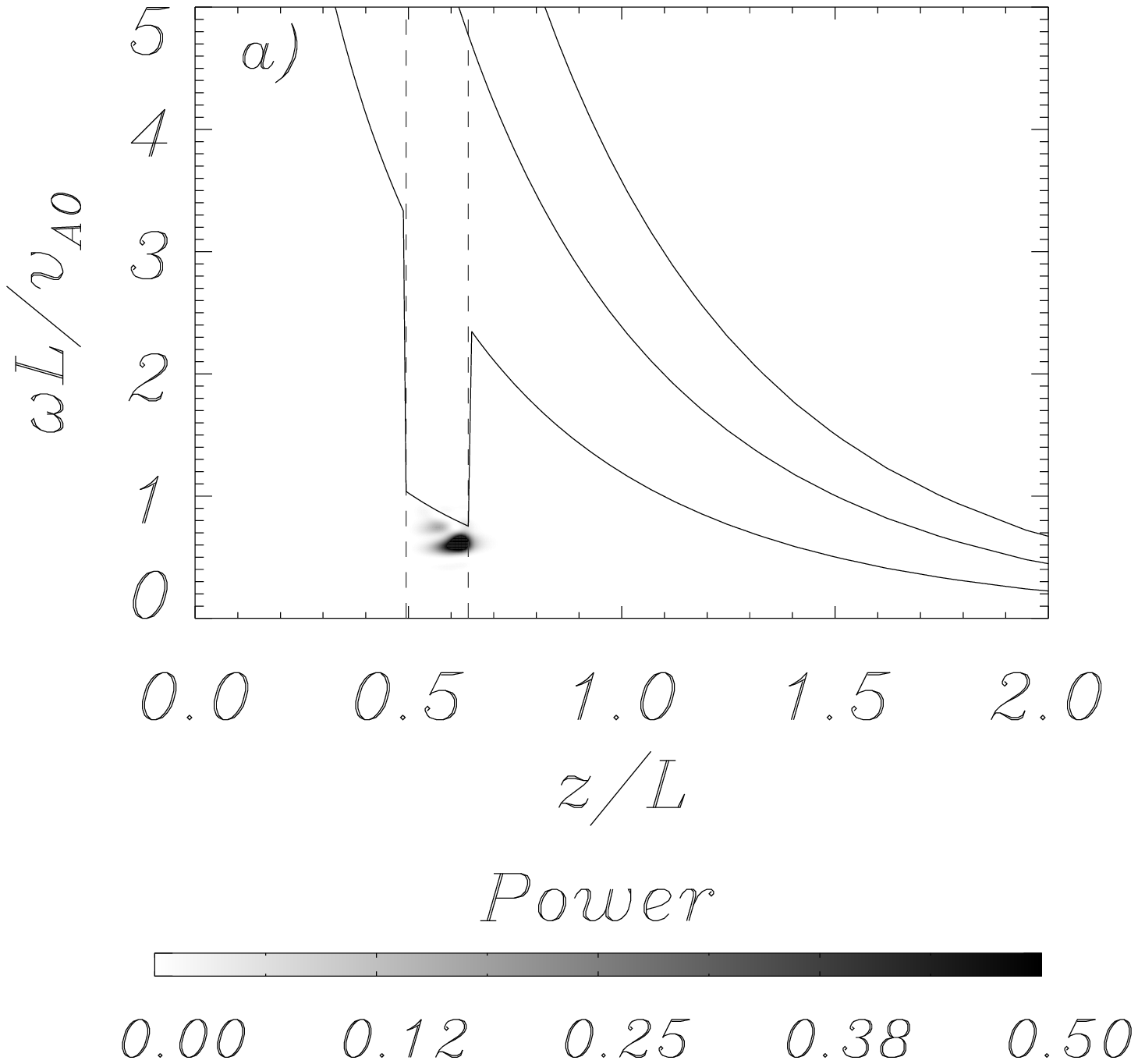}
\includegraphics[width=0.49\textwidth]{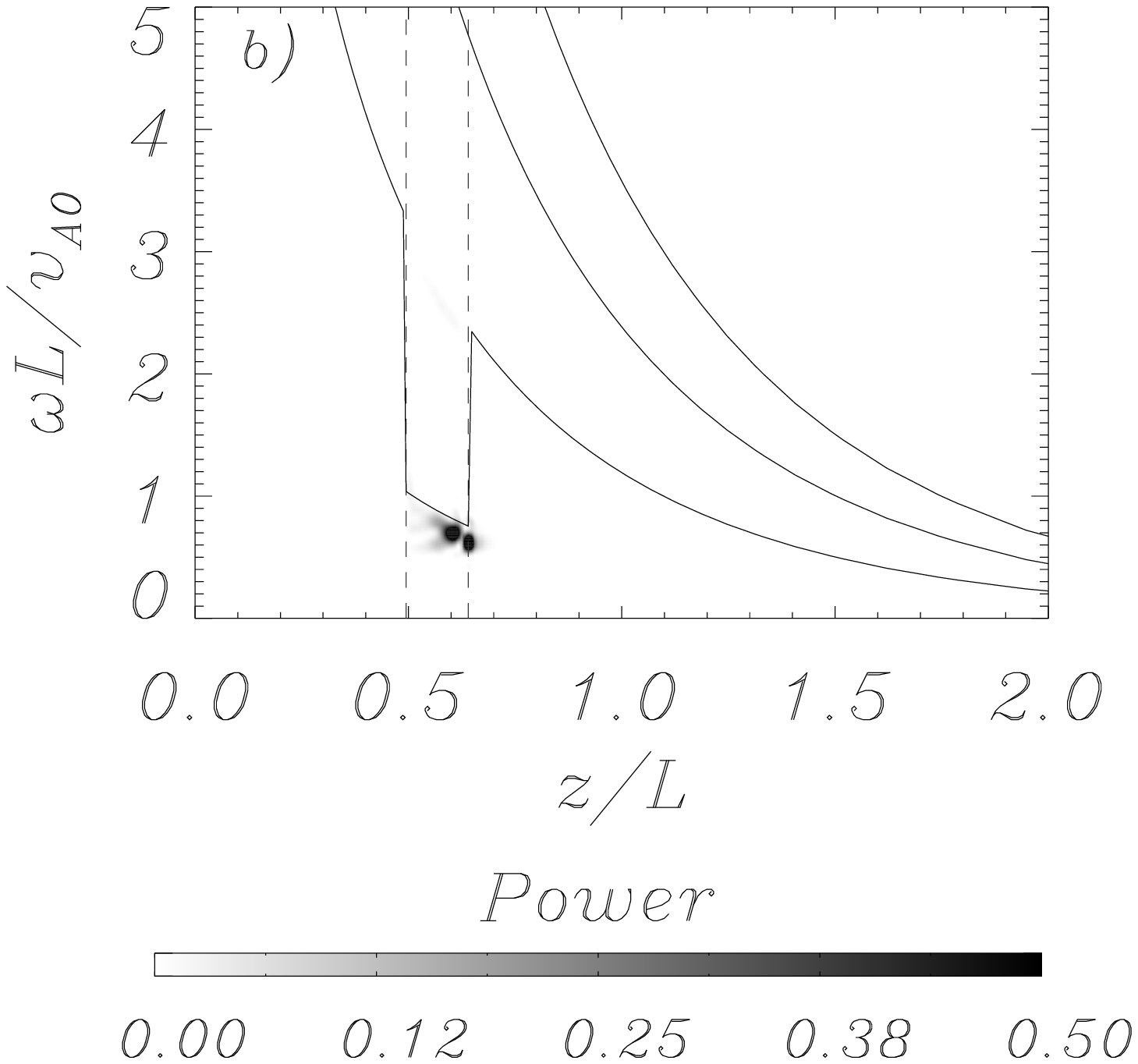}
\caption{Results for $k_yL=60$. Power spectrum of (a) $v_{1n}$ and (b) $v_{1y}$ at $x=0$ as a function of height. The three solid lines are, from bottom to top, the frequency of the fundamental Alfv\'en mode and its first two harmonics. The two dashed lines mark the limits of the coronal loop.}
\label{fig:spectraky60}
\end{center}
\end{figure}

Now, we consider the temporal variation of the normal and perpendicular velocity components along the $z$-axis and compute the power spectra of these two signals, which are then stacked to produce a contour plot of the power as a function of height along the $z$-axis and frequency. Figure~\ref{fig:spectraky60} reveals that the power of $v_{1n}$ and $v_{1y}$ is concentrated inside the loop and that these signals contain two frequencies around $\omega L/v_{A0}=0.6$. A comparison of these graphs with Figures~\ref{fig:vnky5}c and \ref{fig:vyky5}c gives a clear confirmation of the stronger wave confinement achieved by the loop for $k_yL=60$, or in other words, the stronger spatial confinement of the trapped mode of the system. Moreover, since the trapped mode does not extend to large heights, resonant absorption is irrelevant in the present numerical simulation. For resonant absorption to be possible, the trapped mode should have power at a height $z/L\sim 1.5$.

\begin{figure}[h]
\begin{center}
\includegraphics[width=0.45\textwidth]{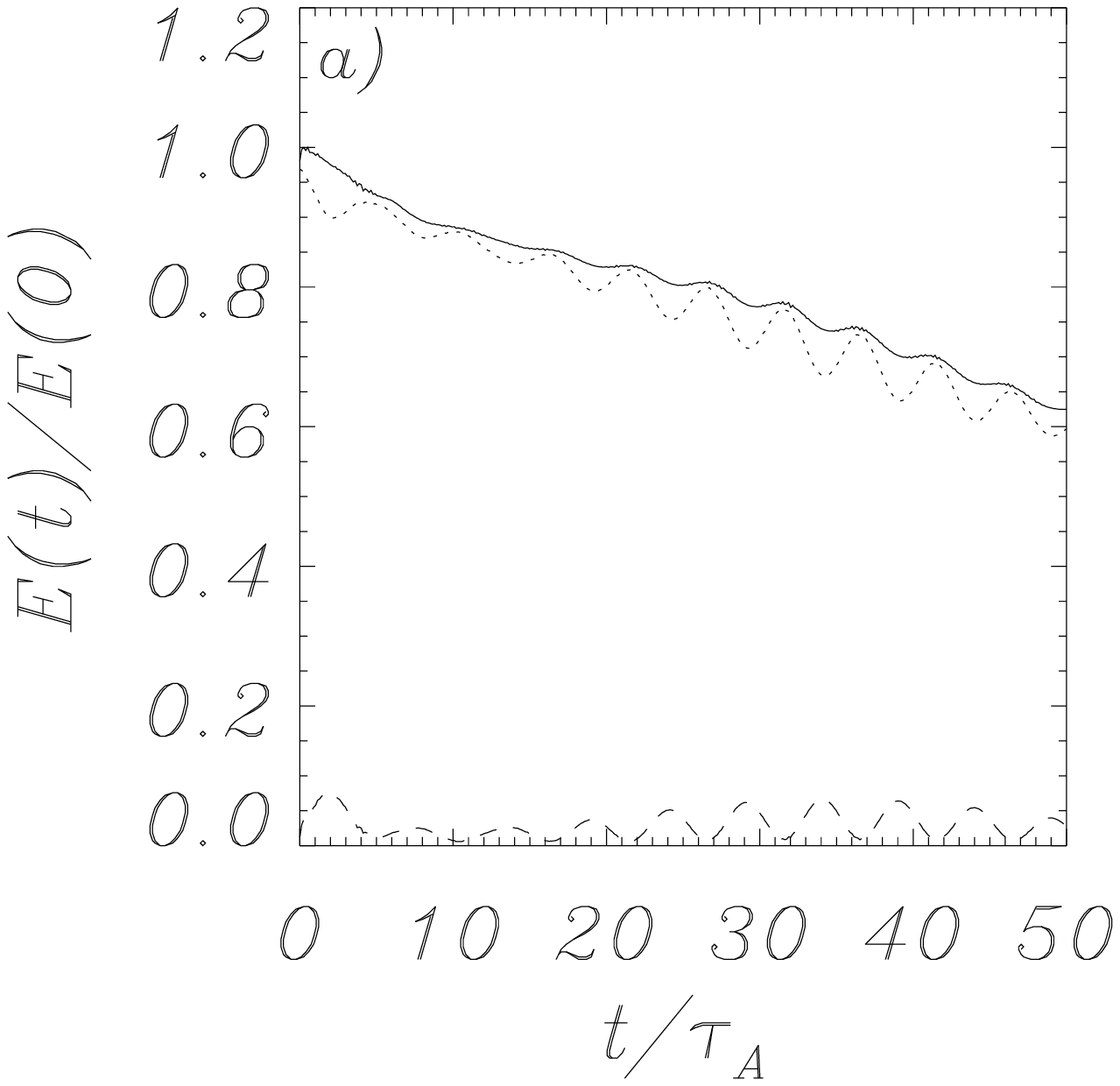}
\includegraphics[width=0.45\textwidth]{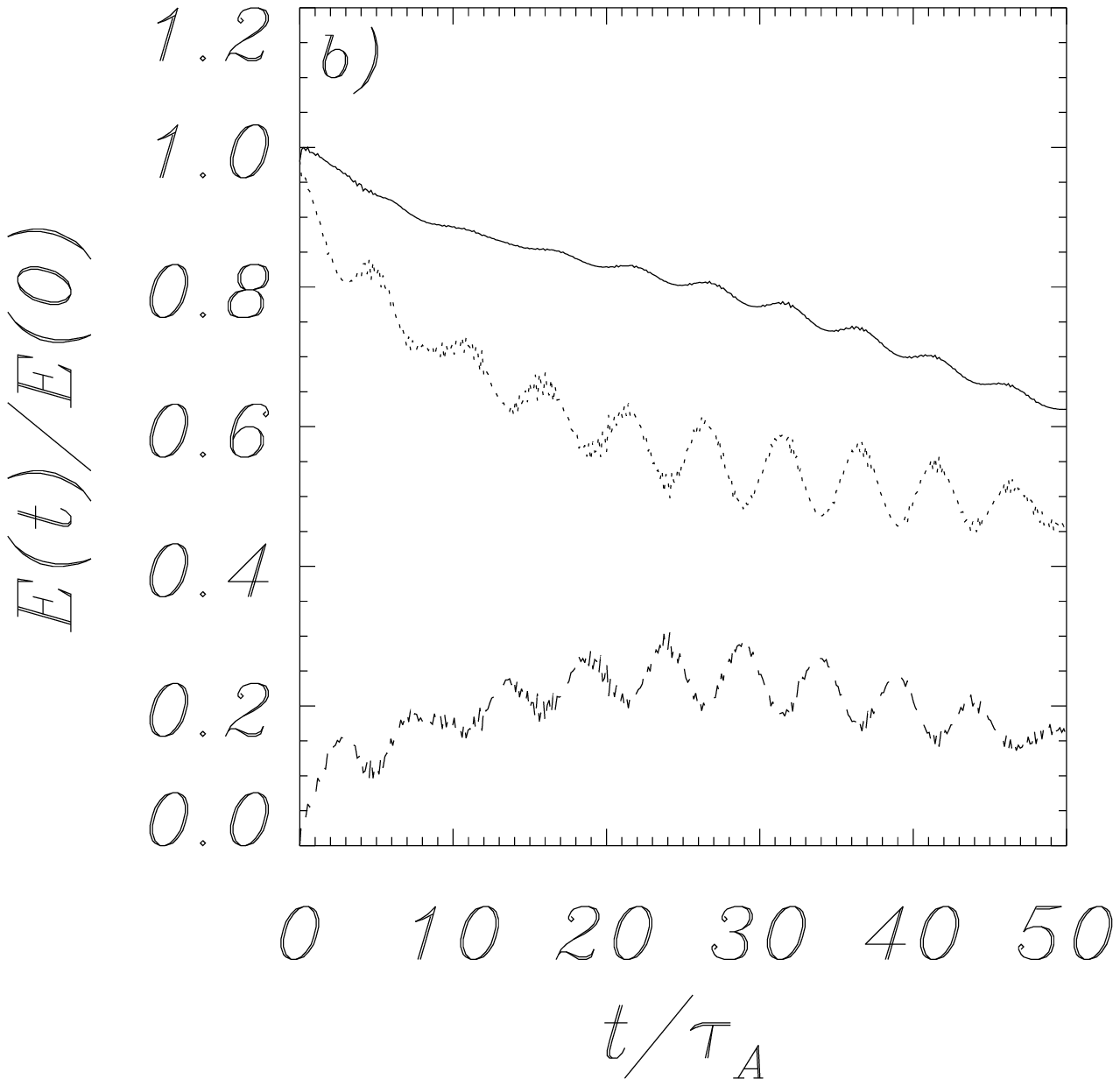}
\caption{Results for $k_yL=60$. (a) Normalized energy density integrated over the whole system (solid line), integrated over the coronal loop (dashed line), and integrated over the region outside the loop (dotted line). (b) Normalized energy density integrated over the whole system contained in all perturbed variables (solid line; same as solid line of panel (a)), in the ``fast'' perturbed variables, $E_{fast}(t)$ (dotted line), and in the Alfv\'enic perturbed variables, $E_{Alfven}(t)$ (dashed line).}
\label{fig:energiesky60}
\end{center}
\end{figure}

We finally pay attention to the energetics. Figure~\ref{fig:energiesky60}a shows that during the whole simulation there is almost no energy outside the loop, in agreement with our previous findings about the wave guiding properties of the loop for a large perpendicular wave number. Regarding the energy transfer from the ``fast'' to the Alfv\'en perturbed variables, Figure~\ref{fig:energiesky60}b indicates that while some energy is transferred from the first to the second one, on the long term the ``fast'' components retain a larger proportion of energy. The total energy of the system presents a decay in time that, unlike to that described in \S~\ref{sect:wave coupling}, is constant during the whole simulation. There are two reasons for this difference: first, the loop is such a good wave guide for transverse oscillations when $k_yL=60$ that very little wave energy is carried out of the system by leaky waves. Then, the initial leaky phase of Figure~\ref{fig:energiesky5} is not found in Figure~\ref{fig:energiesky60}. Second, for $k_yL=5$ resonant absorption brings energy to Alfv\'enic motions at large heights, where the spatial mesh is coarser. This enhances numerical dissipation of Alfv\'en waves, which results in a strong wave damping. For $k_yL=60$, however, all wave energy is retained at small heights, where the spatial mesh is denser. Hence, although numerical dissipation cannot be removed from our numerical experiments, it is greatly reduced.

\subsection{Realistic three-dimensional coronal loop}
Some authors \citep[see for example][]{HY1988} have extended the results of the slab models to the cylindrical geometry by using the equivalent of the azimuthal wave number in Cartesian coordinates. The formula that relates both wave numbers is 

\begin{equation}
k_{y}=\frac{m}{r},
\label{eq:equivwavenumber}
\end{equation}
where $r$ is the loop radius and $m$ is the azimuthal wave number. Since we are interested in vertical oscillations, caused by the kink mode, we take $m=1$. Regarding the loop radius, it is obvious from Figure~\ref{fig:density}a that it changes with height, being largest at the apex and smallest at the feet. Since the physical conditions at the loop top are the most determinant for the kink mode properties, $r$ is taken as the loop half width at the apex. Then we obtain $k_{y}L=16$.

\begin{figure}[h]
\begin{center}
\includegraphics[width=0.3\textwidth]{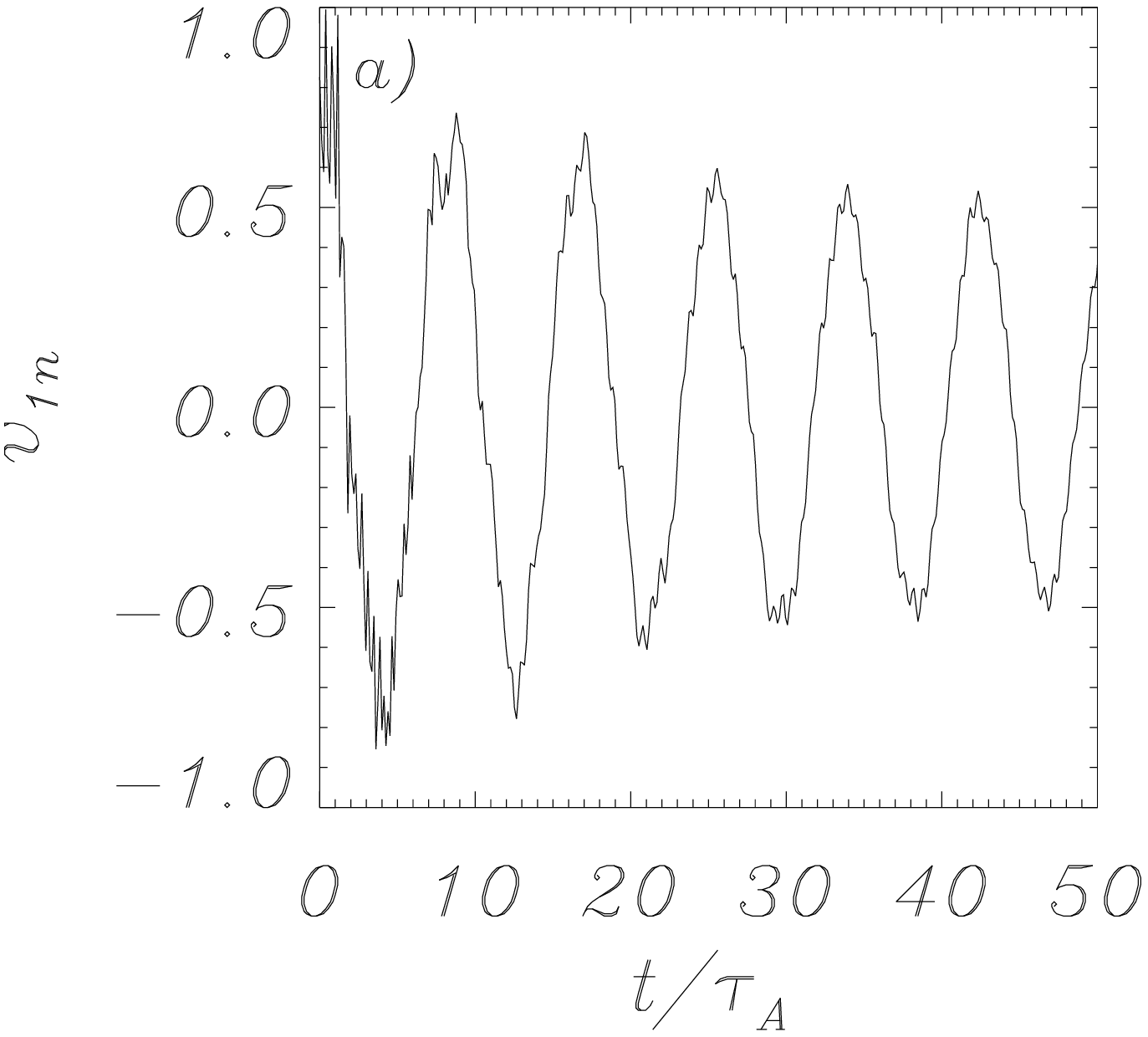}
\includegraphics[width=0.3\textwidth]{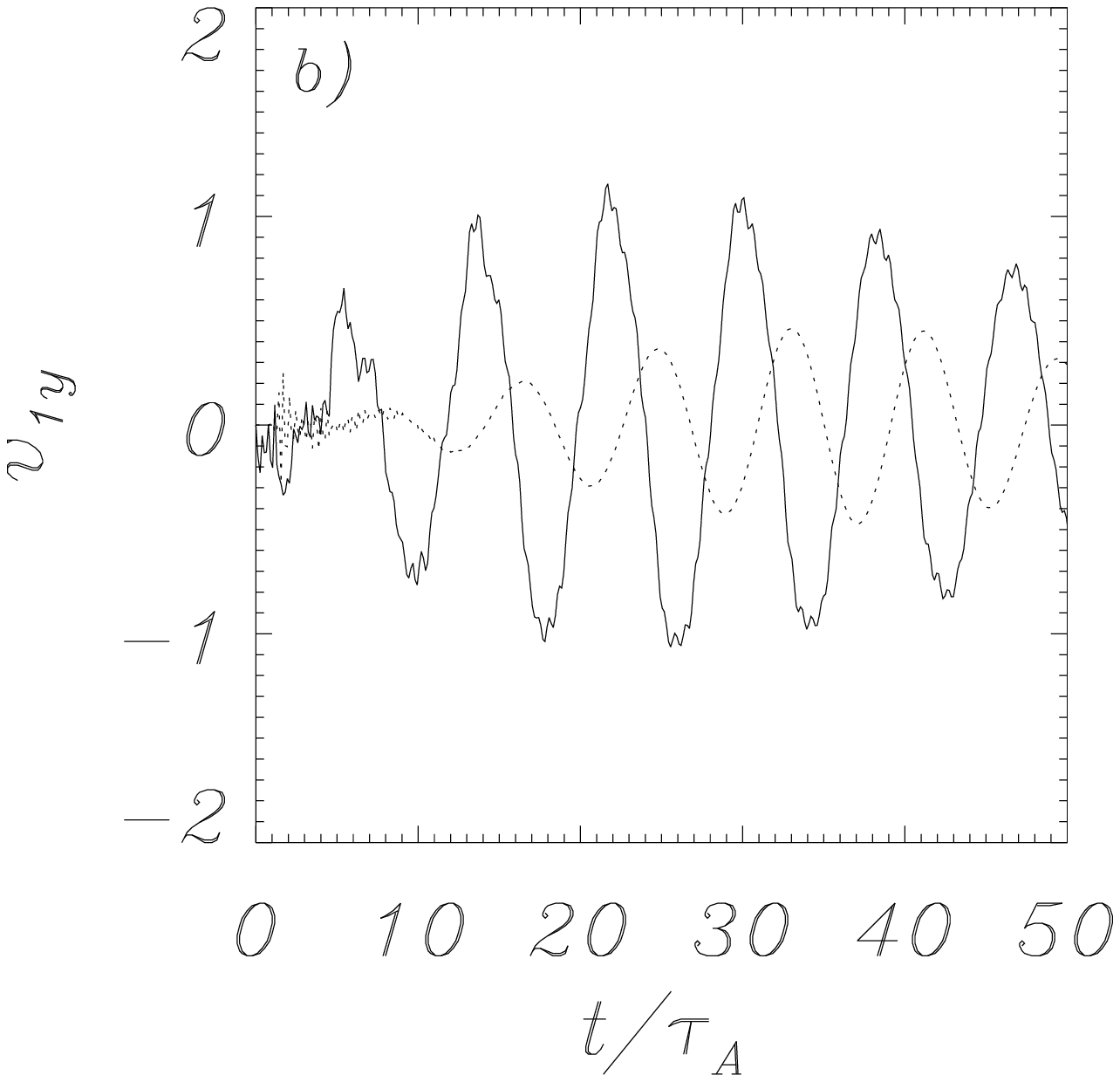} \\
\includegraphics[width=0.3\textwidth]{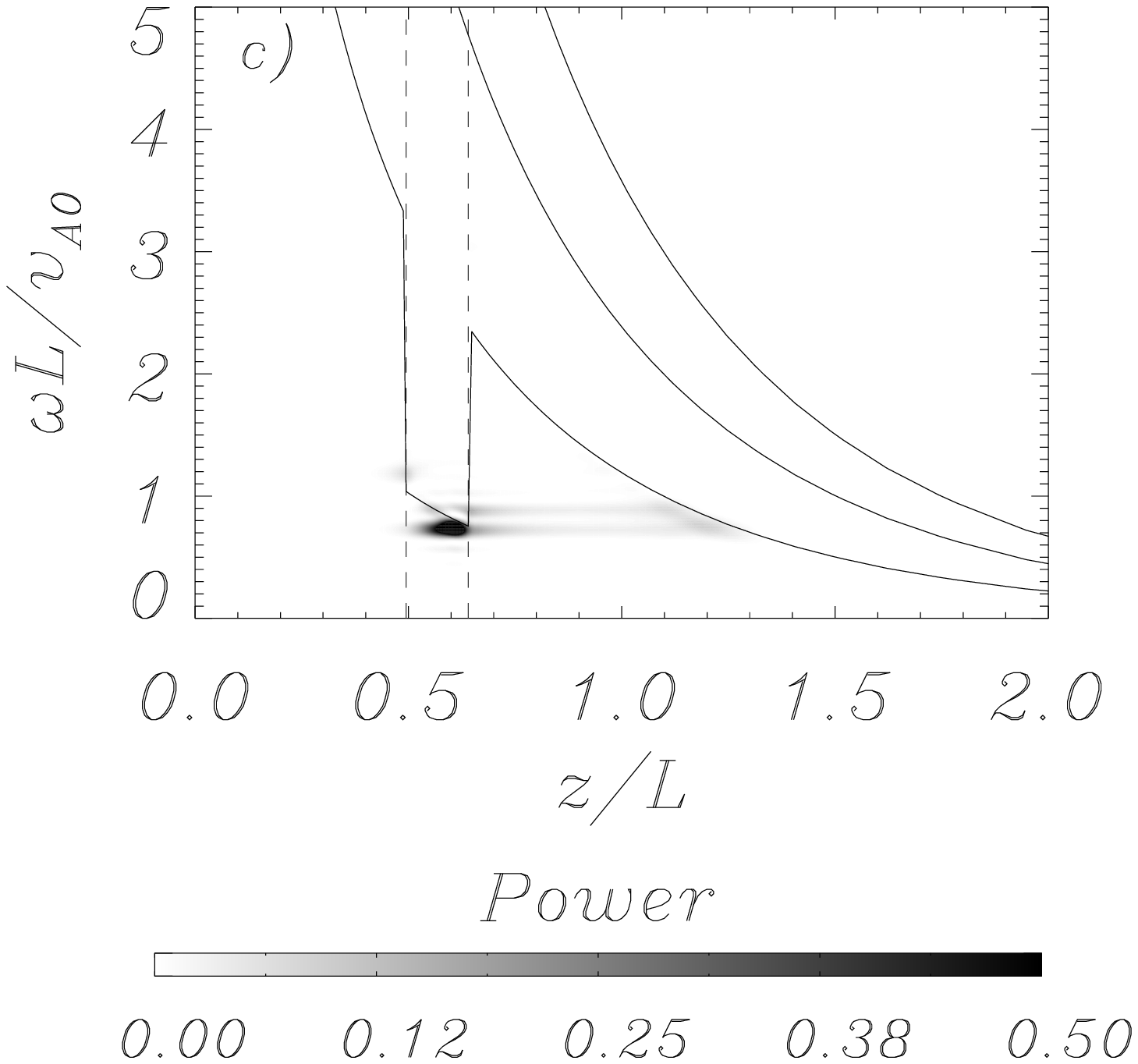}
\includegraphics[width=0.3\textwidth]{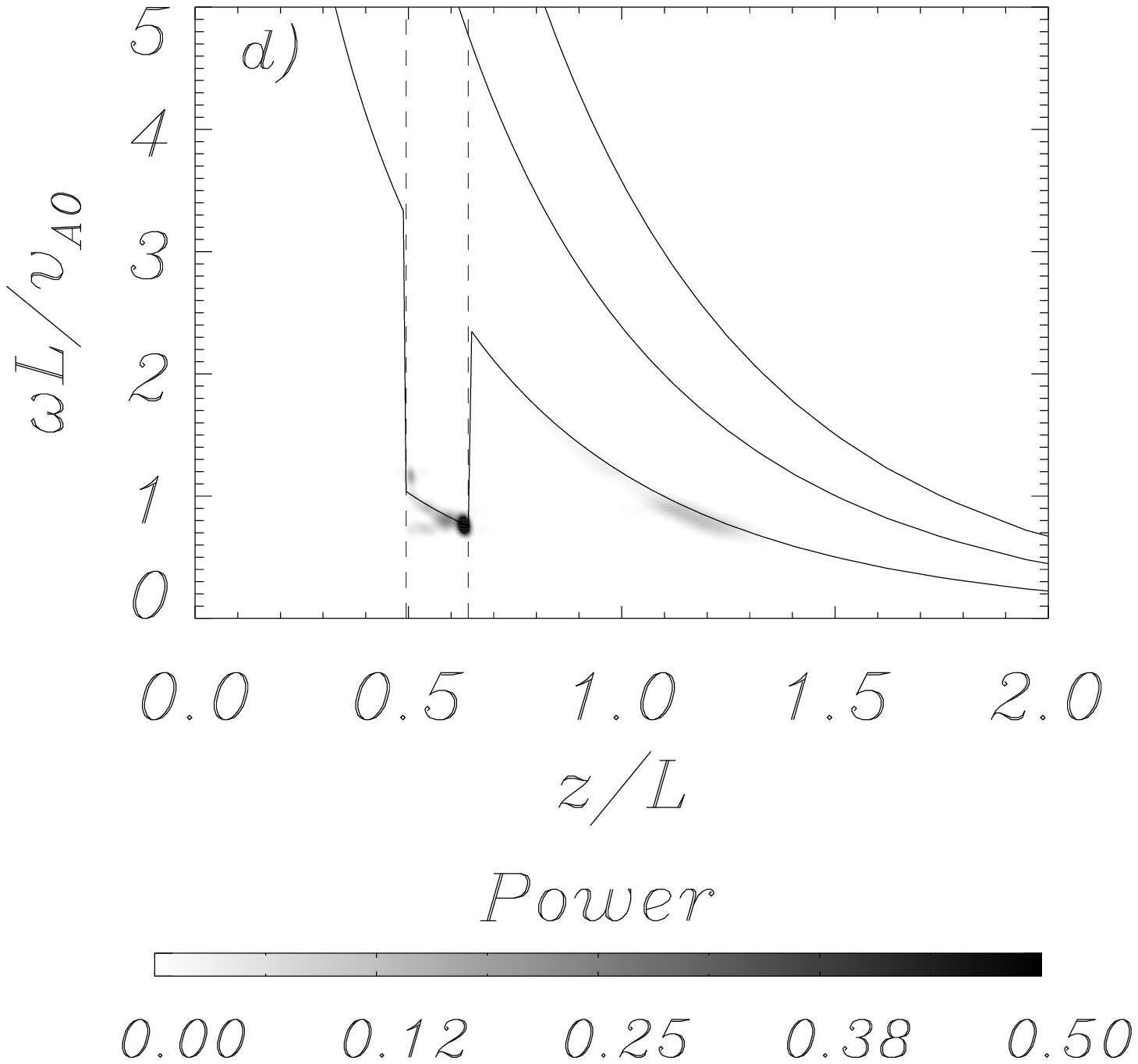}
\caption{Results for $k_yL=16$. (a) Temporal evolution of the $v_{1n}$ velocity component at $x=0$, $z/L=0.61$. (b) Temporal evolution of the $v_{1y}$ velocity component at $x=0$, $z/L=0.61$ (solid line) and at $x=0$, $z/L=1.2$ (dotted line). Power of the normal velocity component for $x=0$ as a function of $z$ and the dimensionless frequency. (d) Power of the perpendicular velocity component for $x=0$ as a function of $z$ and the dimensionless frequency. In panels (c) and (d) the solid lines are the theoretical frequencies of the Alfv\'en continua given by \citet{OBH1993}. }
\label{fig:ampfreqvnvyky16}
\end{center}
\end{figure}

To study the properties of vertical oscillations for this value of $k_y$ we first consider the temporal evolution of $v_{1n}$ at a point inside the loop (see Figure~\ref{fig:ampfreqvnvyky16}a). The periodic behaviour that was found for $k_yL=5$ is also present in this case, although the attenuation rate is smaller now (compare with Figure~\ref{fig:vnky5}a). This suggests a better confinement of the vertical oscillations for $k_yL=16$ and to quantify this effect we compute the power spectra of $v_{1n}$ along the $z$-axis and plot them together as a contour plot, which is presented in Figure~\ref{fig:ampfreqvnvyky16}c. It is clear that two trapped modes, with frequencies $\omega L/v_{A0}=0.74$ and $\omega L/v_{A0}=0.88$, are excited by the initial perturbation and that they are spatially spread along a wide range of heights. Although most of the energy is contained in the fundamental trapped mode and, in particular, inside the loop, the two trapped mode frequencies match the fundamental continuum Alfv\'en frequency around $z/L=1.1-1.3$. This opens the possibility of resonant absorption playing a role in the damping of the trapped modes. To investigate this effect, we plot the perpendicular velocity component at two points along the $z$-axis (Figure~\ref{fig:ampfreqvnvyky16}b). 

The first of these two signals is taken inside the loop, at its top boundary, while the second signal is gathered at one of the resonant positions. Their amplitude aside, these two signals are different in that the first one has an appreciable amplitude just after the initial leaky phase (i.e. for $t/\tau_A\gtrsim 5$), whereas the second one is initially negligible and only after $t/\tau_A\sim 10$ starts to grow, at a rate larger than that of its counterpart inside the loop. To understand these two behaviors we plot the power spectrum of $v_{1y}$ as a function of height and frequency (cf. Figure~\ref{fig:ampfreqvnvyky16}d), which confirms the presence of large power in the transverse velocity component both inside the loop and in the resonant position. The power in the range $z/L=1.1-1.3$ is concentrated around the Alfv\'en continuum frequency and we interpret this as the signature of resonant absorption. On the other hand, inside the loop (i.e. in the range $z/L=0.5-0.64$ there is power at the Alfv\'en continuum frequency but also for $\omega L/v_{A0}\sim 0.74$, that is, at the frequency of the fundamental mode detected in the normal velocity component. We thus conclude that the perpendicular velocity component inside the loop exists both as part of the trapped mode excited by the initial perturbation and by the resonant transfer of energy from this trapped mode to Alfv\'en continuum modes.

\section{Conclusions}
\label{conclusions}
In this paper we have studied the temporal evolution of fast and Alfv\'en waves in a curved coronal loop embedded in a magnetic potential arcade, in order to asses the relevance of three-dimensional propagation of perturbations on the damping of vertical loop oscillations by wave leakage and resonant absorption.

When perpendicular propagation is not included (i.e. when waves are constrained to propagate in the plane of the arcade), a transverse impulsive perturbation produces a combination of leaky modes and the loop is unable to trap energy in the form of vertical kink oscillations. The energy deposited initially in the loop is emitted rapidly to the external medium. This result confirms previous findings in the sense that in a curved coronal loop slab model damping by wave leakage is an efficient mechanism for the attenuation of vertical loop oscillations.

When three dimensional propagation of waves is considered, two new effects are found. First, damping by wave leakage is less efficient and the loop is able to trap part of the energy deposited by the initial disturbance in transverse oscillations. The amount of energy trapped by the structure increases for increasing values of the perpendicular wave number. Second, the inclusion of the perpendicular wave number in a inhomogeneous corona produces the resonant coupling between fast and Alfv\'en modes at those positions where the trapped mode frequency matches that of Alfv\'en waves. This resonant coupling produces the transfer of energy from the fast wave components to Alfv\'enic oscillations. In our model the loop boundary is sharp and so there is no smooth density transition that allows resonant absorption to happen in this position. Hence, this energy transfer does not occur at the loop boundary, but at locations in the external medium, in particular above the coronal loop.

Let us stress that slab models of curved coronal loops in the absence of perpendicular propagation give rise, in general, to damping times by wave leakage shorter than those observed. On the contrary, as we have demonstrated in this work, perpendicular propagation is the clue to obtain damping times compatible with observations.

The authors acknowledge the Spanish Ministry of Science and Technology (MCyT) for the funding provided under projects AYA2006-07637 and AYA2011-22486 and the Government of the Balearic Islands for the funding provided through the Grups Competitius scheme.

\bibliographystyle{../../aastex52/apj}

\end{document}